\documentclass[prd,preprint,tightenlines,floatfix,showpacs,preprintnumbers,nofootinbib]{revtex4}
 \usepackage[dvips,final]{graphicx}
  \usepackage{amssymb}
   \usepackage{amsmath}
    \usepackage{epsfig}
     \usepackage{bm}
      \usepackage{pifont}
\providecommand{\MSbar }{\ensuremath{ \overline{\rm MS} }}
\newcommand{\bit}[1]{\mbox{\boldmath$#1$}}

\def\agoth{\relax\ifmmode{\mathfrak A}\else{${\mathfrak A}${ }}\fi}

\def\muF{\relax\ifmmode\mu_\text{F}^2\else{$\mu_\text{F}^2${ }}\fi}
\def\muR{\relax\ifmmode\mu_\text{R}^2\else{$\mu_\text{R}^2${ }}\fi}
\def\muO{\relax\ifmmode{\mu_{0}^{2}}\else{$\mu_{0}^{2}${ }}\fi}
\def\Mev{\relax\ifmmode{\text{MeV}}\else{MeV{ }}\fi}
\def\MS{$\overline{\text{MS}\vphantom{^1}}${ }}

\def\Li{\relax\ifmmode{\textbf{Li}_{2}}\else{Li$_2${ }}\fi}

\newcommand{\gev}[1]{\relax\ifmmode{\text{GeV}^{#1}}\else{GeV$^{#1}${ }}\fi}

\def\asb{\relax\ifmmode \bar{\alpha}_s\else{$ \bar{\alpha}_s${ }}\fi}
\def\as{\relax\ifmmode \alpha_s\else{$ \alpha_s${ }}\fi}
\def\acal{\relax\ifmmode{\cal A}\else{${\cal A}${ }}\fi}

\def\as{\relax\ifmmode \alpha_s\else{$ \alpha_s${ }}\fi}  
\def\abar{\relax\ifmmode{\bar{a}}\else{$\bar{a}${ }}\fi}  

\newcommand{\be}{\begin{equation}}
\newcommand{\ee}{\end{equation}}

\newcommand{\cc}{\cite}
\newcommand{\ba}{\begin{eqnarray}}
\newcommand{\ea}{\end{eqnarray}}
\newcommand{\bg}{\begin{gather}}
\newcommand{\foma}{\end{gather}}

\newcommand{\noopsort}[1]{}

\newcommand{\vecc}[1]{\mbox{\boldmath $#1$}}

\def\halb{\frac{1}{2}}
\def\e{\epsilon}

\def\w{\omega}

\def\pd{\partial}

\def\F{\Phi}

\def\pb{\bar \psi}

\def\ex{\hbox{e}}
\def\S{\Sigma}
\def\F{\Phi}
\def\<{\langle}
\def\>{\rangle}

\def\a{\alpha}

\def\g{\gamma}  \def\G{\Gamma}
\def\d{\delta}  
\def\l{\lambda}   
\def\s{\sigma}
  
\def\x{\xi}

\def\m{\mu}
\def\n{\nu}
\def\t{\tau}

\def\w{\omega}

\def\({\left(}
\def\[{\left[}
\def\){\right)}
\def\]{\right]}
\def\coth{\hbox{coth}}

\def\pd{\partial}

\def\pa{{\cal P}}

\begin{document}
\thispagestyle{empty}
\date{\today}
\preprint{\hbox{RUB-TPII-01/08}}

\title{Wilson lines and transverse-momentum dependent parton
       distribution functions: A renormalization-group analysis}
\author{I.~O.~Cherednikov\footnote{\sl Alexander von Humboldt Fellow.}}
\email{igorch@theor.jinr.ru}
\affiliation{Institut f\"{u}r Theoretische Physik II,
             Ruhr-Universit\"{a}t Bochum,
             D-44780 Bochum, Germany\\}
\affiliation{Bogoliubov Laboratory of Theoretical Physics, JINR,
             141980 Dubna, Russia\\}
\author{N.~G.~Stefanis}
\email{stefanis@tp2.ruhr-uni-bochum.de}
\affiliation{Institut f\"{u}r Theoretische Physik II,
             Ruhr-Universit\"{a}t Bochum,
             D-44780 Bochum, Germany\\}
\vspace {10mm}
\begin{abstract}
The renormalization-group properties of gauge-invariant
transverse-momentum dependent (TMD) parton distribution
functions (PDF) in QCD are addressed.
We perform an analysis of their leading-order anomalous
dimensions, which are local quantities, making use of the
renormalization properties of contour-dependent composite
operators in QCD.
We argue that attaching individual gauge links with transverse
segments to quark fields in the light-cone gauge, the associated
gauge contours are joined at light-cone infinity through a
cusp-like junction point.
We find that the renormalization effect on the junction point
creates an anomalous dimension which has to be compensated in
order to recover the results in a covariant gauge.
To this end, we include in the definition of the TMD PDF an
additional soft counter term (gauge link) along that cusped
contour.
We show that the eikonal factors entering this counter term are
peculiar to the Mandelstam field formalism and are absent when
one uses a direct gauge contour.
\end{abstract}
\pacs{13.60.Hb,13.85.Hd,13.87.Fh,13.88.+e}
\maketitle


\cleardoublepage

\section{Introduction}
\label{sec:intro}

Theoretical interest in the use of Wilson lines (also termed gauge
links or eikonal phases) has been greatly stimulated recently by
both theoretical investigations and experiments on single spin
asymmetries (SSA)
\cite{ET84,Siv89,Siv90,QS91,Col92,EKT94}---see also \cite{BDR01}
for a quite recent review and further references.
This interest appears in the context of a gauge-invariant formulation
of parton distribution functions (PDF) in terms of hadronic matrix
elements.
Because these matrix elements involve quark--antiquark field operators
at different spacetime points, one has to introduce a path-ordered
gauge link of the form
\be
  [y,x|{\cal C}]
=
  {\cal P} \exp
  \left[-ig\int_{x[{\cal C}]}^{y}dz_{\mu} A_{a}^{\mu}(z) t_{a}
  \right]
\label{eq:Wilson-line}
\ee
that restores gauge invariance, albeit introducing an implicit,
i.e., functional dependence on the contour ${\cal C}$ adopted.
Here, ${\cal C}$ is, in general, an arbitrary path in Minkowski space
and ${\cal P}$ denotes the path-ordering instruction that orders the
Lie-algebra valued gluon fields with the earliest contour point $x$
furthest to the right, whereas the final point $y$ is put furthest to
the left.
[Throughout this work, $t_a$ stands for the generator
$T_{a}^{(\rm F)}$ of the fundamental representation (labelled F)
of color SU$(N_c)$.
Note that there is an implied sum over the color index $a$.]

The concept of gauge links is so pervasive throughout Yang-Mills
theories because it ensures local gauge invariance independent of
the particular dynamical theory.
The concept of using contour-dependent operators in QCD is an old one
and mostly studied in connection with the renormalization of
singularities caused by contour obstructions, like end- or
cross-points, and cusps---see, for instance, in
\cite{Pol79,DV79,Are80,BNS81,CD80,Aoy81,Ste83,IKR85,KR85,KR87,KR91,%
KaKo92,KKS92IW,BKKN93,KaKo94,KKS95,GKKS97,KKSW98,KKS02,Sh00,NSZ00,%
DC2,DC3,DC4} and further references cited therein.
The renewed interest in Wilson lines in SSA and day-to-day applications
is due to the potential breakdown of universality in
transverse-momentum dependent (TMD) parton distribution functions (or
PDF for short) \cite{Col02,BJY02,BMP03,JMY04,JMY04DIS,CM04} and the
ensuing discussion about a process dependence of the gauge link
caused by the color flow of the particular hard-scattering process.
As a consequence, it is argued that there is a change of the overall
sign of the single transverse spin asymmetry from the deeply
inelastic scattering (DIS) to the Drell-Yan (DY) case and, hence,
a breakdown of universality \cite{Col02}.
The content and formulation of this question is coterminous with the
properties of Wilson lines within the purview of specific gauge-fixing
prescriptions and the imposition of boundary conditions at light-cone
infinity.

This brings in questions about the role of final (initial) state
interactions (FSI) between the struck quark and the target spectators
that may yield important effects like shadowing and SSA
\cite{BHMPS02,BHS02DIS,BHS02DY}.
The core issue here is the choice of the gauge adopted for the Wilson
line that has to be inserted to ensure gauge invariance.
For singular gauges, like the light-cone gauge $A^+=0$, the gauge
link vanishes by choice without any restriction put on the gauge
potential ${\bf A}_\perp$ at infinity.
Then, to avoid light-cone singularities at $k^+=0$, specific
boundary conditions on the gluon propagator have to be imposed in
order to exhaust the remaining gauge freedom and recover the
results obtained in non-singular gauges, say, in the Feynman gauge.
It was shown in \cite{JY02}, and further worked out and detailed in
\cite{BJY02}, that the effects of FSI in the light-cone gauge can be
properly taken into account by including a gauge link (an
``eikonalized'' quark line) which involves a path in the transverse
direction.
In covariant, i.e., non-singular, gauges this additional term does not
contribute and, therefore, the modified definition, proposed by
Belitsky, Ji, and Yuan, for the TMD PDF reduces to the correct
gauge-invariant one.
These important findings not withstanding, yet a consistent
gauge-invariant picture for TMD PDFs in the whole phase space is
still incomplete, because the behavior of the gauge contour at
infinite transverse distance is largely arbitrary \cite{JY02}.
The crucial point is---as we have recently shown \cite{CS07}---that
splitting the gauge link and allowing the separated contours to
stretch out to infinity in the transverse configuration space,
induces an additional contribution that cannot be dispensed with by
imposing suitable boundary conditions \cite{BJY02}.
Instead, one has to compensate this new contribution by incorporating
into the definition of the TMD PDF an eikonal factor which provides
a soft counter term in the sense of Collins and Hautmann
\cite{CH99,CH00,Hau07}.

In the present work, we shall investigate these issues from the
point of view of the renormalization group and address parton
distribution functions---integrated and unintegrated---aiming for a
more suitable definition of TMD PDFs.
Our considerations will employ contour-dependent operators and
we will calculate the leading gluon radiative corrections in the
light-cone gauge.
The renormalization of such operators is supremely simple to
deal with when stated in terms of anomalous dimensions because
these quantities are \emph{local} and do not depend on the
length of the gauge contour.
Therefore, they provide a powerful tool to access the renormalization
properties of Wilson lines and take into account those contributions
originating from geometrical obstructions, notably, endpoints,
or sharp bends in the contour, as first pointed out by Polyakov
\cite{Pol79}.
In fact, we will show in more detailed form than in our brief
presentation in \cite{CS07} (see also \cite{CS07spin}) that, adopting
the light-cone gauge, the leading gluon radiative corrections
associated with the transverse gauge link give rise to an anomalous
dimension that exhibits a $\ln p^+$ behavior---characteristic of a
contour with a cusp.
We will fathom out the physics underlying this finding---in particular,
the renormalization effect on the junction point of two individual
gauge contours joined non-smoothly through a cusp.
Moreover, we will present a new definition for the TMD PDF that
(i) reduces to the correct integrated case and
(ii) coincides with the result obtained in the Feynman gauge that is
untainted by contour obstructions (the reason being, we reiterate,
that in this case ${\bf A}_\perp$ vanishes at infinity).

The remainder of the paper is organized as follows.
In the next section, we give a summary of the kinematics used and
sketch the spacetime picture of DIS.
Section \ref{sec:int-PDF} discusses integrated PDFs and their
gauge-invariance and renormalization-group properties.
In Sec.\ \ref{sec:classical}, we reinvigorate our statements on the
transverse gauge link, presented briefly in \cite{CS07}, by a formal
derivation, making use of a ``classical'' current along the lines
of thought described by Jackiw, Kabat, and Ortiz in \cite{JKO91ck}.
The calculation of the leading one-loop anomalous dimension of the
TMD PDF in the light-cone gauge is outlined in Sec.\
\ref{sec:one-loop-LC-gauge}.
Here we also present a generalized factorization rule for gauge
links which takes into account the possibility that the contours
may be joined non-smoothly via a cusp-like junction point.
The same section contains the evaluation of the soft counter term to
supplement the definition of the TMD PDF and its interpretation
as an ``intrinsic Coulomb phase'', in analogy to the phase found
by Jakob and Stefanis \cite{JS91} for QED within a manifestly
gauge-invariant formulation in terms of Mandelstam fields
\cite{Man62,Man68YM}.
In Sec.\ \ref{sec:evolution} we turn our attention to the real-gluon
contributions and the evolution equations, providing a tangible proof
that the integrated PDF obtained from our modified TMD PDF definition
coincides with the correct one with no artefact of the cusped contour
left over.
Section \ref{sec:DY-univ} addresses the application of our approach to
the Drell-Yan case.
Finally, a summary and further discussion of our findings together with
our conclusions is given in Sec.\ \ref{sec:concl}.

\section{Kinematics and spacetime picture of (SI)DIS}
\label{sec:Def-Kin}

\begin{figure}[ht]
 $$\includegraphics[angle=90,width=0.7\textwidth]{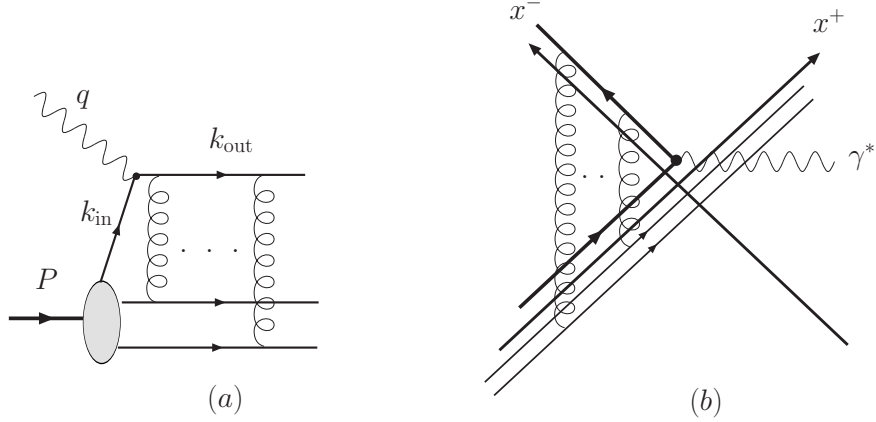}$$
   \vspace{0.0cm} \caption{(a) Schematic Feynman graph for DIS and
                           (b) its spacetime picture at the amplitude
                           level.
                           The thick line denotes the struck quark,
                           whereas the other solid lines along the
                           $x^+$ direction represent spectators, and
                           curly lines mark exchanged gluons.
                           \label{fig:st_dis}}
\end{figure}

In what follows, we employ null-plane coordinates with
\begin{equation}
  P^{\mu} = (P^+,P^-,\vecc P_{\perp}) \ , \
  P^\pm   = (P^0 \pm P^3)/\sqrt{2} \ , \
  P^2     = 2P^+P^- - \vecc P_{\perp}^2 \, ,
\label{eq:hadron-momentum}
\end{equation}
where
$P^+ >0$ is large and the other components
$P^- ,\,  \vecc P_{\perp} >0$ are small.
Moreover, we will visualize the spacetime structure of the chief
hadronic reactions, like DIS, Semi-Inclusive Deep-Inelastic
lepton-nucleon Scattering (SIDIS), etc., as a series of snapshots
on a plane of equal $x^+$ as depicted in Fig.\ \ref{fig:st_dis}.
[For a pedagogical exposition of this graphical method, see, e.g.,
Refs.\ \cite{Sop96,Sop00}, and the review in Ref.\ \cite{CSS89}.]

Let us now fix the kinematics relevant for the cases to be
considered in our work.
We introduce two light-cone vectors
\begin{equation}
  n^{*\m}
=
  \Omega (1, 1, \vecc 0_\perp) \, , \ \
  n^\m = \frac{1}{2\Omega}(1, -1, \vecc 0_\perp)\
\label{eq:lc_vectors}
\end{equation}
with the following properties of their plus/minus light-cone
components
\begin{equation}
  n^{*+}
=
  \sqrt{2} \Omega \, , \
  n^{*-} = 0 \, , \
  n^+ = 0 \ , \
  n^- = \frac{1}{\sqrt{2} \Omega} \, , \ \
  n^* n = 1 \ , \
  (n^*)^2 = n^2 = 0 \ ,
\end{equation}
where $\Omega$ is an arbitrary parameter having the dimension of mass.
Then, the momentum of the initial hadron reads
\begin{equation}
  P^\m
=
  n^{*\m} + \frac{M^2}{2} n^\m \, , \ \
  P^2 = M^2 \ .
\end{equation}
The momentum of the struck quark before being ``measured'' \cite{Sop96}
by the photon is
\begin{equation}
  k_{\rm in}^\m
=
  x P^\m + k_{\perp}^\m \ \
\rightarrow \ \
  k_{\rm in}^+
=
  xP^+ = \sqrt{2}x \Omega
\end{equation}
with
\begin{equation}
  k_{\rm in}^-
=
  \frac{xM^2}{2\sqrt{2} \Omega}  \ , \
  k_{\perp}^\m=(0^+, 0^-, \vecc k_\perp) \ , \
  {{\vecc k}_{\rm in}}_\perp = \vecc k_\perp \, ,
\end{equation}
whereas the off-shell photon has the momentum
\begin{equation}
  q^\m
=
  - x^{\prime} n^{*\m}
  + \frac{Q^2}{2x^{\prime}} n^\m \ \
\rightarrow \ \
  q^+
= -
  \sqrt{2}x^{\prime} \Omega \, , \ \
  q^-
=
  \frac{Q^2}{2\sqrt{2}x^{\prime} \Omega} \ .
\end{equation}
After the interaction with the highly virtual photon the struck
quark acquires the momentum
\begin{equation}
  k_{\rm out}^\m
= (
  k_{\rm in}^\m + q^\m)
= (x - x^{\prime}) n^{*\m}
  + \frac{xx^{\prime} M^2 + Q^2}{2 x^{\prime}} n^\m \ .
\end{equation}

Choosing a specific Lorentz frame upon setting
$\Omega = Q/(2 x^{\prime})$,
corresponds to an almost lightlike hadron that moves along the
plus direction.
In the Bjorken limit $Q^2 \to\infty$, the variables $x^{\prime} $ and
$x$ coincide up to $O(M^2/Q^2)$ terms, so that one gets for the
photon
\begin{equation}
  q^+
= -
  \frac{Q}{\sqrt{2}} \, , \ \
  q^-
=
  \frac{Q}{\sqrt{2}} \ ,
\end{equation}
meaning that the photon moves along the negative $x^3$-direction,
whereas the struck quark (after being probed by the photon) moves
along the minus direction to infinity:
\begin{equation}
  k^+ = 0 \, , \ \
  k^- = \frac{Q^2}{2\sqrt{2} x \Omega} = \frac{Q}{\sqrt{2}} = q^- \ .
\end{equation}
Therefore, in the Bjorken limit, one can estimate the corresponding
conjugated spacetime light-cone coordinates to be
\begin{equation}
  \d x_P^- \sim 1/P^+ \to \mbox{small} \, , \ \
  \d x_P^+ \sim 1/P^- \to \mbox{large} \ ,
\end{equation}
\begin{equation}
  \d x_k^- \sim 1/k^+ \to \mbox{large} \, , \ \
  \d x_k^+
\sim
  1/k^- \to \mbox{small} \ .
\end{equation}
In the next section, the spacetime picture, described above, will be
used to motivate the introduction of the extra transverse gauge link.

\section{Integrated PDF\lowercase{s}: Gauge-invariance and
         renormalization-group properties}
\label{sec:int-PDF}

A well-known example for an integrated PDF is provided by the single
parton distribution of a quark of fractional longitudinal momentum $x$
and flavor $i$ in a fast moving hadron which contains the
nonperturbative physics in DIS:\footnote{Strictly speaking, this is
an unrenormalized quantity.
The renormalization properties of the operators will be considered
shortly in detail.}
\begin{equation}
  f_{i/H}(x)
=
  \frac{1}{2} \int \frac{d\x^-}{2\pi}\
  \ex^{- i k^+ \x^-} \< H(P) |\pb_i
  (\x^-, \vecc 0_\perp) \g^+
  \psi_i(0^-,\vecc 0_\perp) | H (P) \> \ ,
\label{eq:int}
\end{equation}
where the quark momentum is defined by
$k^+ =x P^+$ with $x=Q^2/(2P \cdot q)$.

To give this expression a physical meaning, i.e., elevate it to an
observable, one has to ensure that it is gauge
invariant.\footnote{Otherwise, one would have to define a
``standard'' gauge to be used in all calculations in order to make
them comparable because what is a quark in one gauge is a quark
plus gluons in another.}
To achieve this goal, one usually introduces ad hoc a gauge-link
operator
\begin{equation}
  [y^-, x^-] = {\cal P} \exp
  \left[-i g \int_{x^-}^{y^-} dz^-
        A_{a}^{+}\left(0,z^-, \vecc 0_{\perp}\right) t_{a}
  \right]
\label{eq:link}
\end{equation}
which is path ordered along a lightlike line from the point $x^-$,
where the quark was removed from the hadronic state by the
annihilation operator $\psi_i(0^-,\vecc 0_\perp)$, to the point
$y^-$, where it was recreated by $\pb_i (\x^-, \vecc 0_\perp)$.
The gauge-link operator represents the struck quark as an eikonal
parton \cite{CS82} with fixed color charge $g$, after its
interaction with the highly virtual photon
($0<Q^2=-q^2\gg 1~{\rm GeV}^2$) in the DIS process, as it moves
with (almost) the speed of light along a lightlike line in the
$x^-$ direction.
The spacetime visualization of this process is illustrated in Fig.\
\ref{fig:st_dis}(b).

It should be clear that, ultimately, expressions (\ref{eq:int})
and (\ref{eq:link}) have to be quantized, using, for instance,
functional-derivative techniques.
This means that in employing Eq.\ (\ref{eq:link}) in
Eq.\ (\ref{eq:int}), the gluon potential in the gauge link has to be
Wick contracted with corresponding terms in the interaction Lagrangian,
accompanying the Heisenberg fermion (quark) field operators.
The consequence of this operation is that all terms in the expectation
value in (\ref{eq:int}), proportional to the gauge parameter and
originating from the gluon propagator in a covariant gauge, will
ultimately cancel.
This has been explicitly proved for the quark propagator in
leading-order perturbation theory long ago in \cite{Ste83} (see also
\cite{CD80, Aoy81} and \cite{Kan82} for alternative formulations).
Indeed, it has been shown in these works that after renormalization in
a \MS scheme, the anomalous dimension associated with the ordinary
quark self-energy---which is gauge dependent---gets additional
contributions, stemming from the two endpoints of the gauge link
(coined in \cite{Ste83} the ``connector''),\footnote{These
contributions are generated by singularities at the endpoints of the
line integrals that are dimensionally regularized in $D=4-\epsilon$
dimensions and give $1/\epsilon$ poles.} that exactly cancel its
gauge-parameter term, so that the anomalous dimension of the composite
gauge-independent quark propagator (termed ``hybrid'' in \cite{Ste83})
is indeed free of the gauge parameter.
[Below, the terminology of \cite{Ste83} is adopted.]
This translates into the following sum rule for the anomalous
dimensions
\be
    \gamma_{\rm hybrid}
=
    \gamma_{2q}
  + \gamma_{\rm connector} \, ,
\label{eq:anom-dim-SR}
\ee
where
\be
  \gamma_{2q}
=
    \frac{\alpha_{\rm s} C_{\rm F}  a}{4\pi}
  + {\cal O}(\alpha_{\rm s}^{2}) \, ,
\label{eq:an-dim-quark}
\ee
\be
  \gamma_{\rm connector}
=
    \frac{-\alpha_{\rm s} C_{\rm F} (3+a)}{4\pi}
  + {\cal O}(\alpha_{\rm s}^{2}) \, ,
\label{eq:an-dim-connector}
\ee
and
\be
  \gamma_{\rm hybrid}
=
    \frac{-3 \alpha_{\rm s} C_{\rm F}}{4\pi}
  + {\cal O}(\alpha_{\rm s}^{2})\, ,
\label{eq:an-dim-hybrid}
\ee
with $\alpha_{\rm s}=g^2/4\pi$,
$C_{\rm F}=(N_{\rm c}^2-1)(2N_{\rm c})=4/3$,
and $a$ being the gauge parameter.
Note that we use the same conventions for the definition of the
anomalous dimension as in \cite{Ste83}, i.e., we write
\be
  \gamma_{\rm hybrid}
=
  \frac{\mu}{2}\frac{1}{Z_{\rm hybrid}}\frac{dZ_{\rm hybrid}}{d\mu}
\label{eq:def-an-dim}
\ee
with analogous expressions for the other anomalous dimensions.

The above sum rule (\ref{eq:anom-dim-SR}) is nothing but a
``logarithmic'', i.e., additive, version of the
Ward-Takahashi/Slavnov-Taylor, identities in terms of ratios of the
various renormalization constants of the QCD Lagrangian in the \MS
scheme \cite{DV79,CD80,Aoy81,Ste83}:
\be
  \frac{Z_{\rm connector}Z_{1q}}{Z_{\rm hybrid}}
=
  \frac{Z_{3}}{Z_{1}}
=
  \frac{\tilde{Z}_{3}}{\tilde{Z}_{1}}
=
  1 + \frac{N_{\rm c}\alpha_{\rm s}}{8\pi}(3+a)\frac{1}{\epsilon} \, ,
\label{eq:ST-iden}
\ee
where, $Z_{\rm hybrid}=Z_{\rm connector}Z_{2q}$, with $Z_{2q}$,
$Z_3, \tilde{Z}_{3}$ being, respectively, the renormalization
constants of the quark, the gluon, and the ghost field, whereas
$Z_{1q}, Z_1, \tilde{Z}_1$ are the renormalization constants
pertaining to the quark-gluon-quark vertex, the three-gluon vertex,
and the ghost-gluon-ghost vertex, respectively.\footnote{It is
worth noting that in the covariant gauge $a=-3$, the gauge-contour
divergences cancel among themselves and the Slavnov-Taylor identity
(\ref{eq:ST-iden}) becomes trivially satisfied \cite{CD80,Ste83}.}

Some important comments are here in order:
(i) This leading-order result can be formally proved for smooth
contours in every order of QCD perturbation theory
\cite{GN79,Are80,BNS81}.
(ii) It was shown in \cite{GN79,DV79,CD80,Aoy81} that the connector can
be renormalized by multiplying it with an appropriate renormalization
constant and by replacing in the exponent the strong coupling by its
renormalized version.
Indeed, the renormalization constants $Z_{2q}$ and $Z_{1q}$ do not
depend on the path chosen in any order of $g$
\cite{GN79,Are80,CD80,Aoy81} and the crucial Slavnov-Taylor identities
are satisfied.
This has been explicitly shown in \cite{DV79,Aoy81} up to the order
$g^4$ and to all orders in $g$ in \cite{Are80}.
(iii) The straight line is actually enough for the renormalization
of the connector for any smooth contour \cite{KKS95,Ste96,Ste98}.
The reason is that what matters are only the singularities induced
by the endpoints of the contour, which are multiplicatively
renormalizable using exclusively dimensional regularization
\cite{Ste83}, whereas the specific path itself (for instance, its
length) is irrelevant---provided \emph{no} local obstructions like
cusps and self-intersections are involved.
Such obstructions would induce additional anomalous dimensions because
of discontinuities in the contour slope, as discussed in detail in
\cite{Pol79,CD80,KR85}, that have to be taken into account in the
anomalous-dimensions sum rule.\footnote{Just recently, Pobylitsa
\cite{Pob07} has studied inequalities of a particular class of
anomalous dimensions depending on cusp angles.}
We will show below the key role of a cusped gauge contour in the
eikonalized TMD PDF with a transverse gauge link.

The ``eikonalized'' quark PDF reads
\cite{CS82} (see also \cite{CSS89,CTEQ93})
\begin{equation}
f_{i/a}(x)
=
  \frac{1}{2} \int \frac{d\x^-}{2\pi}
  \ex^{- i k^+ \x^-} \< P |\pb_i
  (\x^-, \vecc 0_\perp)
  \g^+  [\xi^-, 0^-]
  \psi_i(0^-,\vecc 0_\perp) |P\> \ ,
\label{eq:pdf-link}
\end{equation}
and is a manifestly gauge invariant quantity, but has an anomalous
dimension that comprises contributions stemming from the two
endpoints of the lightlike contour ${\cal C}_{\xi^-,0^-}$ (recall the
remarks on the anomalous dimensions given above).
Alternatively, one may be tempted to split the connector $[\xi^-, 0^-]$
into two gauge links that connect the points $0^-$ and $\xi^-$ through
a point at infinity, the aim being to associate each of them with a
quark field operator.
Inserting a complete set of intermediate states, one can then recast
Eq.\ (\ref{eq:pdf-link}) in the following form
\begin{eqnarray}
f_{i/a}^{\rm split}(x)
& = &
  \frac{1}{2} \sum_n \ \int \frac{d\x^-}{2\pi}
  \ex^{- i k^+ \x^-} \< P |\pb_i
  (\x^-, \vecc 0_\perp) [\xi^-, \infty^-]^\dag \ | n \>
  \g^+  \< n | [\infty^-, 0^-]
  \psi_i(0^-,\vecc 0_\perp) |P\>
\nonumber \\
& = &
  \frac{1}{2} \sum_n \ \int \frac{d\x^-}{2\pi}
  \ex^{- i k^+ \x^-}
  \< P |\bar{\Psi}_{i}\left(\xi^-,\vecc 0_\perp|{\cal C}_{1}\right)
       | n \> \g^+
  \< n |\Psi_{i}\left(0^-,\vecc 0_\perp|{\cal C}_{2}\right)
       |P\> \ ,
\label{eq:pdf-link-split}
\end{eqnarray}
where we have introduced the path-dependent Mandelstam fields
\cite{Man62,Man68YM}
\begin{equation}
  \Psi (x| \mathcal{C}_2)
=
 \pa \exp\Bigg[-i g
 \int_{\infty[{\mathcal{C}_2}]}^x d\x_\m A_a^\m (\xi)t_a \Bigg]
 \psi(x) \ ,
\label{eq:mand-psi}
\end{equation}
for the fermion and
\begin{equation}
  \bar \Psi (x | \mathcal{C}_1)
  =  \psi^\dag (x) \pa \exp\Bigg[ i g
  \int_{\infty[{\mathcal{C}_1}]}^x
  d\x_\m  A_a^\m (\xi)t_a \Bigg]\gamma^0  \ .
\label{eq:mand-barpsi}
\end{equation}
for the antifermion.
These field operators represent the struck quark as an eikonal line
\cite{CSS89} along a light-like ray in the minus light-cone direction
while interacting with the gluon field of the hadron, thus mimicking
the motion of a struck quark in a real experiment \cite{Sop96}.

The above two definitions (\ref{eq:int}) and (\ref{eq:pdf-link-split})
are equivalent, because the anomalous dimension of the direct gauge
link $[\xi^-, 0^-]$ is preserved when splitting the contour into two
distinct contours through infinity.
This means, in particular, that the junction point---which is shifted
to infinity---is not creating any anomalous-dimension artefact.
By virtue of the smooth connection of the contours
$\mathcal{C}_1$ and $\mathcal{C}_2$, direct calculation shows that the
renormalization of the junction point $z$ preserves the algebraic
identity
\be
  [x_2,z \ |\ \mathcal{C}_1] \ [z,x_1\ | \ \mathcal{C}_2]
=
  [x_2,x_1\ | \ \mathcal{C}=\mathcal{C}_1\cup \mathcal{C}_2]
\label{eq:link-ident}
\ee
in any order of the coupling \cite{Aoy81}.
This is not trivial because in a local gauge-invariant theory the
factorized gauge link does depend, in general, on the junction point.
However, for purely lightlike smooth contours $\mathcal{C}_1$ and
$\mathcal{C}_2$, the above gauge-link factorization property is
satisfied and the transport of color information by two different
routes does not depend on the junction point (the latter being
``hidden'' at infinity in our case), with complete cancellation
of the contributions at infinity.
This is a well-established property which has been demonstrated
by many authors in the early literature on studies of path-ordered
exponentials.
The analogous situation for non-smooth contours, which are not
purely lightlike, and the generalization of Eq.\
(\ref{eq:link-ident}) will be discussed in the next section.

\begin{figure}[t]
 $$\includegraphics[angle=90,width=0.7\textwidth]{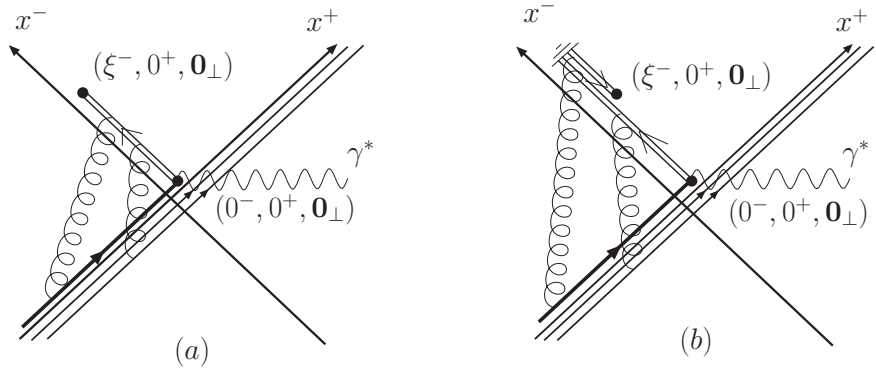}$$
   \vspace{0.0cm} \caption{Space-time picture of the DIS process with
                           a gauge connector (double line) $(a)$ and
                           a split-gauge link (two double lines) at
                           light-cone infinity (b).
                           The infinitely distant parts of the
                           contours, where the behavior of the fields
                           is not specified, are represented by the
                           symbol for ``ground'' (earth) in
                           electricity, introduced in
                           \protect\cite{JS91}.
                           Else, the same designations as in Fig.\
                           \ref{fig:st_dis} are used.
\label{fig:DIS-split}}
\end{figure}

\section{Derivation of the transverse gauge link}
\label{sec:classical}

To substantiate the use of the transverse gauge link and prepare the
ground for the gauge-invariant formulation of unintegrated PDFs, which
bear a full transverse-momentum dependence, let us introduce a Coulomb
source in terms of a ``classical'' current and write
\begin{equation}
  j_\m (y)
  =
  g \int\! d y'_\m \ \d^{(4)} (y - y')\, , \quad
  y'_\m = v_\m \ \t \ ,
\label{eq:current-2}
\end{equation}
which corresponds to a charged point-like particle moving with the
four-velocity $v_\m$ along the straight line $v_\m t$.
The gauge field related to such a current has the form
\begin{equation}
  A^\m (\x)
  =
  \int\! d^4 y \ D^{\m\n}(\x - y) j_\n (y)\ \  ,
\label{eq:source1}
\end{equation}
where $D^{\m\n}$ is the gluon Green's function in an arbitrary
covariant gauge.

Here and in below, we will use two dimensionless light-cone vectors
$n^\pm$
\begin{equation}
  n^\pm = \frac{1}{\sqrt{2}}(1, 0_\perp, \pm 1) \, , \quad
  (n^+)^2 = (n^-)^2 = 0 \, , \quad  (n^+\cdot n^-) = 1 \ ,
\end{equation}
and the metric tensor
\begin{equation}
  g^{\m\n}
  =
  g_T^{\m\n} + {(n^+)}^\m {(n^-)}^\n + {(n^+)}^\n {(n^-)}^\m
  \quad {\rm with} \quad g_T^{\m\n}
  = - \d^{\m\n} \ .
\end{equation}

Appealing to the spacetime structure of this process, illustrated in
Fig.\ \ref{fig:DIS-split}(a), we recast the current in the form
\begin{eqnarray}
  j_\m (y)
& = &
    g \[ n^+_\m \int_{-\infty}^0 \! d\t \ \d^{(4)}(y - n^+ \t)
  + n^-_\m  \int_{0}^\infty \! d\t \ \d^{(4)}(y - n^+ \t)\]
\nonumber \\
& = &
    g \ \d^{(2)} (\vecc y_\perp) \[ n^+_\m \d(y^-)  \int\!
    \frac{dq^-}{2\pi} \frac{\ex^{-iq^- y^+}}{q^- + i0}
    - n^-_\m \d(y^+)  \int\! \frac{dq^+}{2\pi}
    \frac{\ex^{-iq^+ y^-}}{q^+ - i0}\] \ ,
\label{eq:current-3}
\end{eqnarray}
which makes it clear that the first term in this expression
corresponds to a gauge field created by a source moving from minus
infinity to the origin in the plus light-cone direction, before being
struck by the photon, whereas the second term corresponds to a gauge
field being created by a source moving, after the collision, from the
origin to plus infinity along a light-cone ray in the minus light-cone
direction.
Note that the underlying kinematics are those of Sec.\
\ref{sec:Def-Kin}.

To continue, we approximate the gluon Green's function $D^{\m\n}$ by
the free gluon propagator in the light-cone gauge $A^+ = 0$ and obtain
\begin{eqnarray}
  D^{\m\n} (z)
& = &
   \int\! \frac{d^4 k}{(2\pi)^4} \
   {\ex^{- i k \cdot z}} \tilde D^{\m\n}(k)
\nonumber \\
& = &
  - \int\! \frac{d^4 k}{(2\pi)^4} \
    \frac{\ex^{- i k \cdot z}} {k^2+i0}
    \( g^{\m\n}
  - \frac{k^{\mu}(n^-)^{\nu}+k^{\nu}(n^-)^{\mu}}{[k^+]}
    \) \ ,
\label{eq:gluon-prop}
\end{eqnarray}
where $1/[k^+]$ denotes the regularization prescription to handle
the light-cone (pole) singularity at $k^+$.
Taking into account that the $n^-$-part of the current
(\ref{eq:current-3}) does not contribute in the
$A^+ = (A \cdot n^-) = 0 $ gauge,
one is able to obtain the transverse components $\m = i = 1,2$ of the
gauge field; viz.,
\begin{eqnarray}
  A^i_\perp (\x)
& = &
  - g \ n_\n^+ \int\! \frac{d^4 k}{2(2\pi)^4}
   \ {\ex^{- i k \cdot \x}} \ \tilde D^{\m\n} (k)
   \int\! dy^+ dy^- d^2 y_\perp {\ex^{ i k \cdot y}} \d(y^-)
                                \d^{(2)} (\vecc y_\perp)
\nonumber \\
& = &
   - \frac{1}{2}\, g \int\! \frac{dk^+}{2\pi} \
   \frac{\ex^{- i k^+ \x^-}}{[k^+]} \int\! \frac{d^2 k_\perp}{(2\pi)^2}
   \frac{k^i_\perp}{\vecc k_\perp^2} \,
   \ex^{i \bit{\scriptstyle k_\perp}
   \cdot \bit{\scriptstyle \xi_\perp} } \ .
\label{eq:perp_sour1}
\end{eqnarray}
The second, purely transverse, integral over $d^2 \vecc k_\perp$ in
Eq.\ (\ref{eq:perp_sour1}) gets factorized and finally yields
\begin{equation}
    \int\! \frac{d^2 k_\perp}{(2\pi)^2}
    \frac{k^i_\perp}{\vecc k_\perp^2} \,
    \ex^{i\bit{\scriptstyle k_\perp} \cdot \bit{\scriptstyle \xi_\perp}
        }
=
  - \frac{i}{2\pi} \ \nabla^i \ln \l|\vecc  \x_\perp| \ ,
\label{eq:trans-expr}
\end{equation}
while the first integral over $dk^+$ can be performed on account of
the preferred regularization prescription for the pole at $k^+ = 0$.
Note that in Eq.\ (\ref{eq:trans-expr}), $\lambda$ is an auxiliary
infrared (IR) regulator which ultimately drops out from all physical
quantities.
The connection between the regularization procedure in the
momentum-space representation with the behavior of the gauge field at
light-cone infinity can be anticipated from the following expression
\begin{equation}
   \int\! \frac{dk^+}{2\pi} \
   \frac{\ex^{- i k^+ \x^-}}{k^+ \mp i0}
=
   i \kappa \int_0^\infty \! d\l \
   \int\!\frac{dk^+}{2\pi} \ex^{- i \kappa \l (k^+ \mp i0)- i k^+ \x^-}
=
   i \kappa \ \int_0^\infty \! d\l \ \d(\kappa \l + \x^-)\  ,
\end{equation}
where $\kappa = \pm 1$.
Then, one readily finds
\begin{equation}
   \vecc A_\perp (\infty^-; \kappa =  -1) = \frac{g}{4\pi} C_\infty\
   \vecc \nabla \ln \ \l |\vecc \x_\perp| \, ,
\label{eq:ret}
\end{equation}
emphasizing that this expression is dependent on the boundary conditions
via the parameter $C_\infty$.
Obviously, the longitudinal components $A^\pm$ vanish.

Note that the magnitude of $\vecc A_\perp$ in our expression
(\ref{eq:ret}) turns out to be two times smaller than that obtained
in Ref.\ \cc{JKO91ck} (see also \cc{Kov97} and references cited
therein).
The reason for this difference lies in the fact that in our case the
source travels not along the whole plus lightlike axis, but only along
half of it, changing its direction at the origin, as a result of its
collision with the hard photon, and in agreement with the physical
picture of the process.

Let us continue by supplying within the same approach the gauge field
in a general covariant gauge, characterized by the gauge parameter $a$,
and using the gluon propagator
\begin{equation}
  \tilde D^{\m\n} (k) = - \frac{1}{k^2 + i0}
  \left[ g^{\m\n} - (1-a) \frac{k^\m k^\n}{k^2 + i0} \right] \ .
\end{equation}
Starting from Eq.\ (\ref{eq:source1}), one gets after some simple
algebraic manipulations
\begin{equation}
   \vecc A^{\prime}_\perp
= 0 \, ,
\quad
   A^{\prime -}
=  0 \, ,
\quad
   A^{\prime +} (\x)
=
   - \frac{g}{4\pi} \d(\x^-) \ln \ \l |\vecc \x_\perp| \ ,
\label{eq:A-prime}
\end{equation}
where fields in a covariant gauge are marked by a prime accent in
order to distinguish them from those in the light-cone gauge
[cf.\ Eq.\ (9) in Ref.\ \cc{JKO91ck}].
Notice that the $a-$dependent terms, which are proportional to
$\sim k^-$ under the $d^4 k$- integral, do not contribute by virtue
of the delta-function $\d(k^-)$.
Next we give the (singular) gauge transformation which connects these
two representations in the light-cone gauge and in covariant gauges:
\begin{equation}
  A_\m^{\rm LC}
= A^{\prime}_\m + \pd_\m \ \phi \, , \quad
  \phi (\x)
=
  - \int_{-\infty}^{\x^{-}}\!  d\x^{\prime -}
    A^{\prime +} (\x^{\prime -}) \ .
\label{eq:gauge_trans}
\end{equation}

Now we are ready to discuss the origin of the transverse contribution
in Mandelstam's gauge-invariant formalism in the light-cone gauge.
First, note that the analogous expression to the Mandelstam field
(\ref{eq:mand-psi}) in a covariant gauge reads
\begin{equation}
  \Psi_{\rm cov} (\x ; n^+)
=
  \pa \exp\left[-i g
  \int_{\x^-}^{\infty^-} \! dz^- A_{\rm cov}^+ (z^-, \vecc \xi_\perp)
          \right]
  \psi_{\rm cov}(\x^-,\vecc \x_\perp)  \  , 
\label{eq:mand_lc}
\end{equation}
where the gauge field $A_{\rm cov}^+$ differs from the special case,
$A^{\prime +}$, given by Eq.\ (\ref{eq:A-prime}).
Second, performing a regular gauge transformation
\begin{equation}
  U(x^-)
=
  \exp{\left(-ig\int_{}^{x^-}dz^- A^+\right)}
\end{equation}
in the light-cone gauge on both sides of (\ref{eq:mand_lc}), one can
eliminate the Wilson-line integral in the phase.
However, the regular transformation $U(x^-)$ does not exhaust the gauge
freedom in the light-cone gauge completely and is, therefore,
insufficient to trivialize the interaction of the struck quark with
the gluon field of the spectators.
More explicitly, a residual singular transformation
$U_{\rm sing} (\infty^-, \vecc \x_\perp)$
is still allowed and one realizes that the singular gauge
transformation (\ref{eq:gauge_trans}) reflects exactly this
remaining gauge freedom.
Carrying out this additional gauge transformation, one generates an
additional phase that is now associated with the quark field itself;
viz.,
\begin{eqnarray}
  \psi(\x^-,\vecc \x_\perp)_{\widehat{\rm LC}}
& = &
  U_{\rm sing} (\infty^-, \vecc \x_\perp) \psi_{\rm LC}(\x^-,\vecc \x_\perp)
\nonumber \\
& = &
  \[ 1 - i g \int_{-\infty^-}^{\infty^-}\!
   dz^- A_{\rm source}'^+ (z^-, \vecc \xi_\perp)
   + O(g^2)
          \]
          \psi_{\rm LC}(\x^-,\vecc \x_\perp) \ ,
\label{eq:sing-gauge-transfo}
\end{eqnarray}
which is now completely gauge-fixed, and hence, represents a quark
with a fixed color charge.
[This is indicated by a wide hat over the label LC which abbreviates
`light cone'.]
Finally, by taking into account expression (\ref{eq:ret}), one finds
that the quark wave function in the light-cone gauge acquires a phase
that may formally be written as
\begin{equation}
   \psi(\x^-,\vecc \x_\perp)_{\widehat{\rm LC}}
=  \[1 + i g\int_{\x_\perp}^{\infty_\perp}\!
    d\vecc z_\perp \vecc A_{\rm source}^{\rm LC}
   (\infty^-, \vecc z_\perp) + O(g^2) \]
   \psi_{\rm LC}(\x^-,\vecc \x_\perp) \ .
\end{equation}
The above arguments make it clear that a complete gauge fixing can be
achieved in (\ref{eq:sing-gauge-transfo}) by inserting the additional
singular gauge transformation $U_{\rm sing}$ which contains the
cross-talk effects of the struck parton with the light-cone source.
As a result, taking the product of two (local) quark field operators
in the fixed light-cone gauge differs from what one finds in a
covariant gauge.
This difference is encapsulated in two phase factors so that one gets
\begin{eqnarray}
   \[  \bar \psi(\x^-, \vecc \x_\perp) \gamma^+
   \psi  (0^-, \vecc 0_\perp) \]_{\widehat{\rm LC}}
  & = &
  \bar \psi_{\rm LC} (\x^-, \vecc \x_\perp)\pa
  \exp\[+  ig \int_{\x_\perp}^{\infty_\perp}\!
  d\vecc z_\perp \vecc A_{\rm source}^{\rm LC}
  (\infty^-, \vecc z_\perp) \] \gamma^+ \
\nonumber \\
&& \times
  \pa \exp\[-  i g\int_{0_\perp}^{\infty_\perp}\!
  d\vecc z_\perp \vecc A_{\rm source}^{\rm LC}
  (\infty^-, \vecc z_\perp) \]  \psi_{\rm LC} (0^-, \vecc 0_\perp) \ .
\nonumber \\ \label{eq:lc_perp}
\end{eqnarray}
In the next section, we will use the results obtained above in order
to demonstrate the role of the transverse link in the restoration of
the prescription-independence of the anomalous dimension of the TMD PDF.
Specifically, the explicit expression for the transverse gauge field at
infinity, Eq.\ (\ref{eq:perp_sour1}), will be used in the diagrammatic
calculations of the gluon radiative corrections pertaining to the
anomalous dimension of the TMD PDF.

\section{Calculation of the one-loop anomalous dimension
         of the TMD PDF in the light-cone gauge}
\label{sec:one-loop-LC-gauge}

We have stressed before the importance of the renormalization effect on
the junction point of the decomposed transverse contours at infinity.
In this section we will explain exactly what this means in mathematical
detail.
We will prove that the factorization of the gauge link into factors,
each associated with a distinct contour starting (ending) at light-cone
infinity, has to be modified to include an additional phase factor which
accounts for the cusp anomalous dimension induced by the junction point
of these decomposed contours.

\subsection{Definitions}
\label{subsec:def}

Our starting point is the operator definition of the (unpolarized)
TMD distribution of a quark with momentum
$k_\m = (k^+, k^-, \vecc k_\perp)$
in a quark with momentum $p_\m = (p^+, p^-, \vecc 0_\perp)$:
\begin{equation}
\begin{split}
   f_{q/q}(x, \mbox{\boldmath$k_\perp$})
 = {}&
   \frac{1}{2}
   \int \frac{d\xi^- d^2
   \mbox{\boldmath$\xi_\perp$}}{2\pi (2\pi)^2}\,
   \ex^{- i k^+ \x^-}
   {\rm e}^{+ i \bit{\scriptstyle k_\perp}
\cdot \bit{\scriptstyle \xi_\perp}}
   \left\langle  q(p) |\bar \psi (\xi^-, \vecc \xi_\perp)
   [\xi^-, \mbox{\boldmath$\xi_\perp$};
   \infty^-, \mbox{\boldmath$\xi_\perp$}]^\dagger \right.\\
& \times [\infty^-, \mbox{\boldmath$\xi_\perp$};
   \left. \infty^-, \mbox{\boldmath$\infty_\perp$}]^\dagger 
   \gamma^+[\infty^-, \mbox{\boldmath$\infty_\perp$};
   \infty^-, \mbox{\boldmath$0_\perp$}]
   [\infty^-, \mbox{\boldmath$0_\perp$};0^-, \mbox{\boldmath$0_\perp$}] \right. \\
& \times \left. \psi (0^-,\mbox{\boldmath$0_\perp$}) |q(p)\right\rangle \
   |_{\xi^+ =0}\ ,
\label{eq:tmd_definition}
\end{split}
\end{equation}
where the lightlike and transverse gauge links are defined,
respectively, by
\begin{equation}
\begin{split}
[ \infty^-, \mbox{\boldmath$z_\perp$}; z^-, \mbox{\boldmath$z_\perp$}]
\equiv {}&
 {\cal P} \exp \left[
                     i g \int_0^\infty d\tau \ n_{\mu}^- \
                      A_{a}^{\mu}t^{a} (z + n^- \tau)
               \right] \ ,  \\
[ \infty^-, \mbox{\boldmath$\infty_\perp$};
 \infty^-, \mbox{\boldmath$\xi_\perp$}]
\equiv {}&
 {\cal P} \exp \left[
                     i g \int_0^\infty d\tau \ \mbox{\boldmath$l$}
                     \cdot \mbox{\boldmath$A$}_{a} t^{a}
                     (\mbox{\boldmath$\xi_\perp$}
                     + \mbox{\boldmath$l$}\tau)
               \right] \ .
\label{eq:lines_definition}
\end{split}
\end{equation}
Let us emphasize that in the definition of the transverse gauge link
the contour is defined in terms of the two-dimensional vector
$\vecc l$ which is absolutely arbitrary.
We will show explicitly that this arbitrariness does not affect the
local properties of the gauge link---in particular the anomalous
dimension.

Employing the light-cone axial gauge
\begin{equation}
   A^+
=
  (A \cdot n^-) = 0 \ , \ {(n^-)}^2= 0 \ ,
\label{eq:axial-gauge}
\end{equation}
the gluon propagator has an additional pole, related to the plus
light-cone component of the gluon momentum, and reads
\begin{equation}
   D_{\mu\nu}^{\rm LC} (q)
=
   \frac{-i}{q^2 - \lambda^2 + i0} \Big( g_{\mu\nu}
  -\frac{q_\mu n^-_\nu + q_\nu n^-_\mu}{[q^+]}\Big) \ .
\end{equation}
To give this expression a mathematical meaning, we apply the following
pole prescription
\begin{equation}
  \frac{1}{[q^+]}\Bigg|_{\rm Ret/Adv}
=
  \frac{1}{q^+ \pm i \eta} \ \ \ , \ \ \
  \frac{1}{[q^+]}\Bigg|_{\rm PV}
=
  \frac{1}{2} \[ \frac{1}{q^+ + i \eta} + \frac{1}{q^+ - i \eta} \] \ ,
\label{eq:pole-prescription}
\end{equation}
where $\eta$ has the dimension of mass, to be kept small but finite,
and where we used the abbreviations Ret for retarded, Adv for advanced,
and PV for the principal value.\footnote{We remark that another
possible prescription---the so-called Mandelstam-Leibbrandt pole
prescription \cite{Man82,Lei83,Lei87}---is outside the scope of the
present investigation, though we will make some related comments in
connection with the anomalous dimension in Eq.\ (\ref{eq:gamma_2}).}
In what follows, we regularize collinear poles by means of the quark
virtuality $p^2 < 0$, whereas IR singularities are regularized by an
auxiliary gluon mass $\lambda$ which is put at the end back to zero.
The described regularization procedure works well in the one-loop
order, while at a higher loop order one may need to apply more
sophisticated methods.

In the tree approximation, where the gauge links are equal to unity and
the quark-gluon interactions vanish, one trivially gets
\begin{equation}
\begin{split}
  f_{q/q}^{(0)} (x, \mbox{\boldmath$\xi_\perp$})
= {} &
  \frac{1}{2}
  \int \frac{d\x^- d^2\mbox{\boldmath$\xi_\perp$}}{2\pi (2\pi)^2}
  \ex^{- i k^+ \x^- + i \bit{\scriptstyle k_\perp}
\cdot
  \bit{\scriptstyle \xi_\perp}}
  \< p |\pb (\x^-, \mbox{\boldmath$\xi_\perp$})
  \g^+
  \psi (0^-,\mbox{\boldmath$0_\perp$}) | p \>
\\
=
  {} &
  \d (p^+ -x p^+) \d^{(2)}(\vecc k_\perp)
  \frac{1}{2} \bar u(p) \g^+ u (p) \ ,
\end{split}
\end{equation}
where $u(p)$ denotes a quark spinor and summation over spin indices
is tacitly assumed.
Moreover, we use the short-hand notation
\begin{equation}
   \frac{\bar u (p) \g^+ u (p)}{2 p^+}
   \equiv
   \phi_0(p) \ ,
\end{equation}
which implies
\begin{equation}
   f_{q/q}^{(0)} (x, \mbox{\boldmath$\xi_\perp$})
   =
   \delta (1-x) \d^{ (2) } (\vecc k_\perp) \ \phi_0(p) \ .
\end{equation}

\subsection{One-loop calculation}
\label{subsec:one-loop-calc}

\begin{figure}
\centering
\includegraphics[scale=0.7,angle=90]{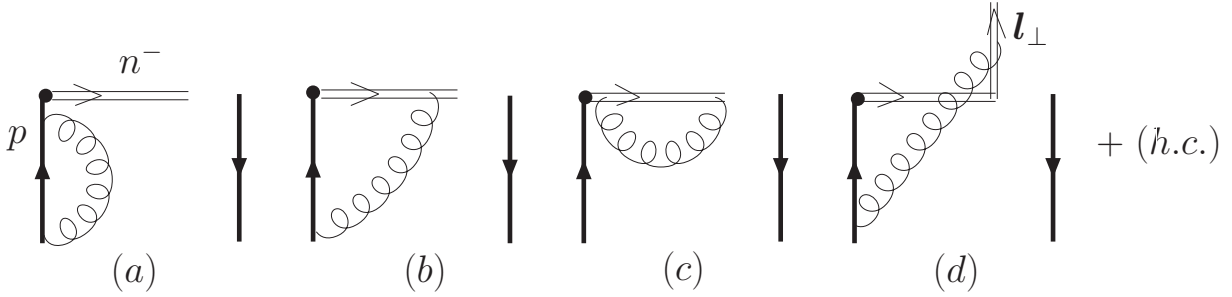}~~
\caption{One-loop gluon contributions (curly lines) to the
         UV-divergences of the TMD PDF in a general covariant gauge.
         Double lines denote gauge links.
         Diagrams (b) and (c) are absent in the light-cone gauge.
         The omitted Hermitian conjugate diagrams are symbolically
         abbreviated by $(h.c)$.
\label{fig:se_gluon}}
\end{figure}

Our objective here is to discuss the leading-order (LO) $g^2$
quark-gluon interactions, stemming, one one hand, from the standard
QCD vertices, and, on the other hand, from the interactions of the
quarks with the gauge links.
Because in the TMD case the distance between the quark fields is
spacelike, i.e., $\xi_\m \xi^\m = - \vecc \x_\perp^2 \neq 0$,
the UV-divergent contributions arise only due to virtual gluon
corrections.
Therefore, in LO (alias, at the one-gluon exchange level), diagrams
(a), (b), and (c) contribute in a covariant gauge (see
Fig.\ \ref{fig:se_gluon}), while in the light-cone gauge, only
diagrams $(a)$ and $(d)$ give non-vanishing contributions---with the
latter diagram being associated with the transverse gauge link.

The Hermitian-conjugate ($h.c.$ for short) contributions (omitted in
Fig.\ \ref{fig:se_gluon}) are generated by the corresponding
``mirror'' diagrams.
In what follows, we consider first the ``left'' set of diagrams
(in order to show explicitly how the transverse gauge link comes into
play) and then we take the sum ``left + right'' which gives the total
contribution to the TMD PDF.

These diagrams yield contributions proportional to delta-functions:
$$
  f_{q/q}^{\rm 1-loop}(x, \vecc k_\perp; \mu, \eta)
=
  \d(1-x)\d^{(2)}(\vecc k_\perp)\phi_0(p)\Sigma (p, \a_s, \mu, \eta)\ ,
$$
where ${\S}^{\rm 1-loop}$ results from the diagrams shown in Fig.\
\ref{fig:se_gluon}.
The quark self-energy diagram $(a)$ gives (in dimensional
regularization with $ \w = 4 - 2 \e $)
\begin{equation}
   {\S}^{(a)} (p, \a_s ; \mu, \eta, \e)
= -g^2 C_{\rm F} \m^{2\e}\ \int\! \frac{d^\w q}{(2\pi)^\w}
   \frac{\g_\m (\hat p - \hat q) \g_\n}{(p-q)^2
   (q^2 - \lambda^2 + i0)}\, d^{\m\n}_{\rm LC}(q)
   \frac{i\hat p}{p^2} 
\label{eq:self-energ}
\end{equation}
with
\begin{equation}
  d^{\m\n}_{\rm LC} (q)
=
  g^{\m\n} - \frac{q^\m (n^{-})^\n}{[q^+]} - \frac{q^\n (n^{-})^\m}{[q^+]}\ ,
\label{eq:gluon_pr}
\end{equation}
where the dependence on the auxiliary mass scale $\eta$ is ``hidden''
in the pole prescription $[q^+]$ and
$
  \hat a \equiv (\g \cdot a) \ .
$

The $g^{\m\n}$-proportional term gives a ``Feynman''-like contribution,
namely,
\begin{equation}
   {\S}_{\rm Feynman}^{(a)} (p, \a_s, \mu, \e )
=
  -g^2 C_{\rm F} \m^{2\e}\ \int\! \frac{d^\w q}{(2\pi)^\w} \
  \frac{\g_\m (\hat p - \hat q) \g^\m}{(p-q)^2 (q^2 - \lambda^2) }
  \frac{i \hat p}{p^2}
\label{eq:fey1}
\end{equation}
and generates no extra light-cone singularities.
After carrying out the momentum integral, one gets
\begin{equation}
   {\S}_{\rm Feynman}^{(a)} (p, \a_s, \mu , \e )
=
  -\frac{\a_s}{4\pi} C_{\rm F} \G(\e)
   \(-4\pi\frac{\m^2}{p^2}\)^\e (1-\e) \int_0^1\! dx
   \[x(1-x)
     \(1 - \frac{\lambda^2}{x p^2} \)
   \]^{-\e} \ .
\label{eq:fey2}
\end{equation}
Performing the remaining integral, one finally finds
\begin{eqnarray}
{\S}_{\rm Feynman}^{(a)} (p, \a_s,  \m , \e )
 = &&
  \!\!  - \frac{\a_s}{4\pi} C_{\rm F} \G(\e) (1-\e)
    \(-4\pi\frac{\m^2}{p^2}\)^\e
\nonumber \\
&& \times
   \[ 1 +
     \e \( 2 + \frac{\lambda^2}{p^2}
           \ln \frac{\lambda^2 - p^2}{\lambda^2}
          -\ln \frac{p^2 - \lambda^2 }{p^2}
        \) + O(\e^2)
   \] \ .
\label{eq:fey3}
\end{eqnarray}
Note that the ''mirror'' diagram gives precisely the same contribution,
doubling this result.

\subsection{Evaluation of the pole-prescription-dependent
            contributions}
\label{subsec:pole-prescr}

The calculation of the $[q^+]$-dependent part
\begin{equation}
   {\S}^{(a)}_{\rm pole} (p, \a_s, \m , \eta  ; \e )
=
   g^2 C_{\rm F} \m^{2\e}\!
   \int\! \frac{d^\w q}{(2\pi)^\w}
   \frac{1}{(p-q)^2 (q^2 - \lambda^2 )}
   \(
      \frac{\hat q (\hat p - \hat q) \gamma^+ }{[q^+]}
     +\frac{\gamma^+ (\hat p - \hat q) \hat q}{[q^+]}
   \)
     \frac{i \hat p}{p^2} 
\label{eq:pole1}
\end{equation}
is more demanding, owing to the presence of light-cone
singularities, and we will consider its evaluation in detail.
After some simple transformations of the numerator, we find
\begin{equation}
\begin{split}
    {\S}^{(a)}_{\rm pole}
=
    g^2 C_{\rm F} \m^{2\e}\ \int\! \frac{d^\w q}{(2\pi)^\w}
    \[ ( \hat p \g_\m \gamma^+ + \gamma^+ \g_\m \hat p)
    \frac{(p - q)^\m}{(p-q)^2}
  -2\gamma^+ \] \frac{1}{(q^2 - \lambda^2 ) [q^+]}
    \frac{i \hat p}{p^2}\ .
\label{eq:pole2}
\end{split}
\end{equation}
One observes that the integral
$\int\! \frac{d^\w q}{(q^2 - \lambda^2 ) [q^+]}$
vanishes (which is true for the considered pole prescriptions but not
for the Mandelstam-Leibbrandt one), while the rest can be recast in
the form
\be
  {\S}^{(a)}_{\rm pole}
=
  g^2 C_{\rm F} \m^{2\e}
  (\hat p \g_\m \gamma^+ + \gamma^+ \g_\m \hat p)
  \[ p^\m \s_1(p) + n^\m \s_2(p) \] \frac{i \hat p}{p^2}\ ,
\label{eq:pole3}
\ee
where
\begin{equation}
  \s_1(p)
=
  \frac{i}{(4\pi)^{\w/2}} \frac{\G(\e)}{(-p^2)^\e} \int_0^1\!
  dx\frac{(1-x)}{[xp^+]}
  \[x (1-x) \( 1 - \frac{\lambda^2 }{x p^2}\) \]^{- \e}
\label{eq:sigma-1}
\end{equation}
bearing in mind that by virtue of
$\gamma^+ \gamma^+ = {(n^-)}^2 = 0$
the term $\s_2 (p)$ does not contribute.
Thus, one has
\begin{eqnarray}
  \left[ {\S}_{\rm Feynman}^{(a)}
        +{\S}^{(a)}_{\rm pole}
  \right]\left(p, \a_s , \m , \eta, \e\right)
= \!\!\!&&
   \frac{\a_s}{4\pi} C_{\rm F}
   \(-4\pi\frac{\m^2}{p^2}\)^\e \ \G(\e)
   \left\{\rule{0in}{3.0ex} (1-\e) \right.
\nonumber \\
~~~~~\times \!\!\!\!\!\! && \left.
   \[ 1 + \e \( 2 + \frac{\lambda^2}{p^2}
      \ln \frac{\lambda^2 - p^2}{\lambda^2}
     -\ln \frac{p^2 - \lambda^2 }{p^2}
             \)
   \] \right. \nonumber \\
&&
\left. \!\!\!\!\!\! - \frac{2\gamma^+ \hat p}{p^+} \int_0^1\!
       dx\frac{(1-x)}{[x]}
   \left\{ 1 - \e \ln \[x (1-x) \( 1 - \frac{\lambda^2 }{x p^2}
                                \)
                      \]
   \right\}
\right. \nonumber \\
~~~~~+ \!\!\!\!\!\!&&  \left. \rule{0in}{3.0ex}  O(\e^2)
    \right\} \ .
\label{eq:fey+pole}
\end{eqnarray}

In order to evaluate the integral $\int {dx (1-x)}/{[x]}$, one has to
use a specific pole prescription for $[x]$.
Let us consider three possible prescriptions: Advanced, Retarded and
Principal Value:
\begin{equation}
   \frac{1}{[x]}_{\rm Ret} =  \frac{1}{x + i\bar\eta}\ , \quad\quad
   \frac{1}{[x]}_{\rm Adv} =  \frac{1}{x - i\bar\eta}\ , \quad\quad
   \frac{1}{[x]}_{\rm PV}  =  \frac{1}{2}\(\frac{1}{x+i\bar\eta}
                             +\frac{1}{x - i\bar\eta}\)
\end{equation}
using temporarily for convenience the short-hand notation
$\bar\eta = \eta/p^+$.
In the limit of small $\bar\eta$, we keep only logarithmic terms and
omit any powers of $\bar\eta$.
The UV-divergent part (in the \MSbar-scheme) then reads
\begin{equation}
   {\S}_{\rm UV}^{(a)}
=
  -\frac{\a_s}{4\pi} C_{\rm F}\frac{1}{\e}
   \[1  - \ln 4 \pi + \gamma_E  - \frac{2\gamma^+ \hat p}{p^+}
   \( 1 + \ln \frac{\eta}{p^+} - \frac{i\pi}{2} - i \pi \ C_\infty \)
   \] \ ,
\label{eq:s_div1}
\end{equation}
where the numerical constant $C_\infty$ depends on the pole
prescription according to (cf.\ \cite{BJY02})
\begin{equation}
  C_\infty
=
  \left\{
  \begin{array}{ll}
  & \ \ 0  \ , \ {\rm Advanced}    \\
  & - 1 \ , \ {\rm Retarded}  \\
  & - \frac{1}{2} \ , \ {\rm Principal~~Value~.}
  \
  \end{array} \right.
\label{eq:c_inf}
\end{equation}
One the other hand, the finite part of the pole-prescription dependent
gluon radiative corrections is
\begin{eqnarray}
  {\S}_{\rm finite}^{(a)}{(p, \a_s,\m, \eta, \e )}
= \!\!\! &&
  - \frac{\a_s}{4\pi} C_{\rm F}
  \left(
         1 + \ln \frac{\m^2}{p^2}
           + \frac{\lambda^2}{p^2}\ln \frac{\lambda^2 - p^2}{\lambda^2}
           - \ln \frac{p^2 - \lambda^2 }{p^2} \right.
\nonumber \\
&& \left.  - \frac{2\gamma^+ \hat p}{p^+}
     \left\{
       \(
         1 + \ln \frac{\eta}{p^+} - \frac{i\pi}{2} - i \pi \ C_{\infty}
       \)
         \ln \frac{\m^2}{p^2} \right. \right.
\nonumber \\
&&  + \left. \left. \int_{0}^{1} dx \frac{(1-x)}{[x]}
             \ln \left[
                       x(1-x) \left(1-\frac{\lambda^2}{x p^2}\right)
                 \right]
     \right\}
   \right) \ .
\label{eq:finite-pp}
\end{eqnarray}

Evaluating this UV-finite integral (\ref{eq:finite-pp}) by setting
$\lambda^2 = 0$ (which is justified given that this integral is IR
finite and the magnitude of the IR regulator is irrelevant), we obtain
\begin{equation}
  \int_0^1\! dx \frac{(1-x)}{x \mp i\eta } \ln \left[x (1-x) \right]
=
  2 + \(1 \mp i\eta \) \[ {\rm Li}_2 \( \frac{1}{i\eta}\) -
   {\rm Li}_2 \( \frac{1}{1 - i\eta}\) \] \ .
\label{eq:integral}
\end{equation}
As a result, the complete prescription-dependent finite part for
$\lambda^2 = 0$ becomes
\begin{eqnarray}
  {\S}_{\rm finite}^{(a)}{(p, \a_s, \m, \eta, \e )}
=\!\!\! &&
  - \frac{\a_s}{4\pi} C_{\rm F}
  \left( 1 + \ln \frac{\m^2}{p^2}
   - \frac{2\gamma^+ \hat p}{p^+}
   \left\{ \( 1 + \ln\frac{\eta}{p^+} - \frac{i\pi}{2} - i \pi C_\infty
           \)
   \ln \frac{\m^2}{p^2} \right. \right.
\nonumber \\
&& \left. \left. +
2 + \(1 \mp i\eta \) \[ {\rm Li}_2 \( \frac{1}{\pm i\eta}\) -
   {\rm Li}_2 \( \frac{1}{1 \mp i\eta}\) \]
   \right\} \right) \ .
   \label{eq:finite_fin}
\end{eqnarray}
A key remark here is that any dependence on the pole prescription in
Eqs.\ (\ref{eq:s_div1}) and (\ref{eq:finite-pp}) is cancelled by
taking into account analogous contributions originating from the
transverse gauge link as depicted in Fig.\ \ref{fig:se_gluon}(d).

\subsection{Contribution of the transverse gauge link at light-cone
            infinity}
\label{subsec:trans-gauge-link}

The issue at stake in this subsection is the proof of the cancellation
of all pole-prescription dependent terms in dealing with the
gauge-invariant formulation of the TMD PDF.
To be more specific, we focus on the effects related to the
interactions with the gluon field of the transverse gauge link
\begin{equation}
 \pa \exp\[+ i g \int_0^\infty\! d\t \vecc l_{\perp}
 \cdot \vecc A_{\perp} (\infty^-, 0^+; 
 \vecc l_{\perp} \t + \vecc \x_\perp)\] \
 \pa \exp\[- i g \int_0^\infty\! d\t \vecc l_{\perp}
 \cdot \vecc A_{\perp} (\infty^-, 0^+; \vecc l_{\perp} \t )\] \ .
\label{eq:trans-gauge-link}
\end{equation}
and show that the contribution of the diagram \ref{fig:se_gluon}(d)
cancels out all terms proportional to the numerical factor $C_\infty$
in Eqs.\ (\ref{eq:s_div1}) and (\ref{eq:finite-pp}) (or equivalently
(\ref{eq:finite_fin})).
Before we proceed, note that, as it is obvious from Eq.\
(\ref{eq:s_div1}), the UV-divergent part depends not only on the pole
prescription but also on the logarithmic $p^+$-term.
The effects related to this latter dependence will be considered
subsequently.

In Sec.\ \ref{sec:classical}, we have worked out the transverse
components of the gluon field in the light-cone gauge and found
Eq.\ (\ref{eq:perp_sour1}).
This expression can be further evaluated to read
\begin{equation}
  \vecc A_{\perp} (\infty^-, 0^+; \vecc l_{\perp} \t)
=
  \int\! \frac{dq^+}{2\pi}\ \ex^{- i q^+ \infty^-} \!\!\!
  \int\! \frac{d^2 q_\perp}{(2\pi)^2}
  \ex^{i \bit{\scriptstyle q_{\perp}}\cdot \bit{\scriptstyle l_{\perp}}}
  \vecc A_{\perp} (q) \ ,
\label{eq:trans_1}
\end{equation}
finally assuming the form
\begin{equation}
  \int_0^\infty\! d\t \vecc l_{\perp} 
  \cdot \vecc A_{\perp} (\infty^-, 0^+; \vecc l_{\perp} \t)
=
  \int\! \frac{dq^+}{2\pi}
  \ex^{- i q^+ \infty^-}\int\! \frac{d^2 q_\perp}{(2\pi)^2} \
  \vecc l_{\perp} \cdot \vecc A_{\perp} (q) 
  \frac{i}{(\vecc q_{\perp} \cdot \vecc l_{\perp}) + i0} \ .
\label{eq:trans_2}
\end{equation}
Consider now the free gluon propagator resulting from the correlation
between longitudinal and transverse gluons; viz.,
\begin{equation}
  \< A^\m (q) \vecc A_{\perp}^i (q^{\prime}) \>
=
  - \frac{q^i n^{-\m}}{( q^2 - \lambda^2 ) [q^+]}
  (-i) (2\pi)^4 \d^{(4)} (q + q^{\prime}) \ ,
\label{eq:trans_3}
\end{equation}
and use the relation
\begin{equation}
  \frac{\ex^{- i q^+ \infty^-}}{[q^+]}
= 2\pi i C_\infty \d(q^+)
\label{eq:trans_4}
\end{equation}
to find the contribution of the diagram in Fig.\ \ref{fig:se_gluon}(d):
\begin{eqnarray}
   {\S}_{\perp}^{(d)} (p, \m , g; \e)
& = &
   g^2 C_{\rm F} \m^{2\e} 2\pi i C_\infty \int\!
   \frac{d^\w q}{(2\pi)^\w} \d (q^+) \
   \frac{\gamma^+ (\hat p - \hat q) }{(p-q)^2 (q^2 - \l^2) }
\nonumber \\
& = &
   i C_\infty \a_s C_{\rm F} \(-4\pi\frac{\m^2}{p^2}\)^\e \G(\e)
   \frac{\gamma^+ \hat p}{2 p^+}
   \int_0^1\! \frac{(1-x) \delta(x)}{[x(1-x)]^\e}
   \(1- \frac{\lambda^2}{xp^2}\)^{-\e}
 \ .
\label{eq:perp}
\end{eqnarray}
One sees explicitly that by virtue of the relation
\begin{equation}
  \frac{1}{[x]}
=
  \lim_{\eta \to 0} \frac{1}{x \pm i\eta}
=
  {\rm PV} \frac{1}{x} \mp i\pi \delta(x) \ ,
\label{eq:pole-prescr}
\end{equation}
the transverse-gauge link contribution (\ref{eq:perp}) exactly cancels
the dependence on the pole prescription in both the UV-divergent part
${\S}_{\rm UV}^{(a)}$ and in the finite part
${\S}_{\rm finite}^{(a)}$.
To show this explicitly, we collect all pole-prescription dependent
terms of diagram (a) in Fig.\ \ref{fig:se_gluon} and add to them the
contribution from the transverse gauge link, i.e., Eq.\
(\ref{eq:perp}).
Then, we have
\begin{eqnarray}
    \lim_{\eta \to 0} \int_0^1\!
    dx (1-x) && \!\!\!\!\! \!\!\!\!\!\[
    \frac{1+ C_\infty }{x - i \eta } -
    \frac{C_\infty }{x + i \eta }
  - i 2 \pi C_\infty \delta (x) \] \ln \(x (1-x) \)
\nonumber \\
\!\!\! & = &
    \lim_{\eta \to 0}
    \int_0^1\! dx
    \frac{1 - x}{x - i \eta } \ln \(x (1-x) \)
\end{eqnarray}
which establishes the independence of the result on the parameter
$C_\infty$---the latter encoding the adopted pole-prescription.
As a result, the complete UV-divergent part of the TMD PDF
$f_{q/q}(x, \mbox{\boldmath$k_\perp$})$ is
\begin{eqnarray}
  {\S}^{(a+d)}_{\rm UV} (p, \m, \a_s ; \e)
& = &
  - \frac{\a_s}{\pi}\ C_{\rm F} \frac{1}{\e}
    \[\frac{1}{4}- \frac{ \gamma^+ \hat p}{2 p^+}
  \( 1 +  \ln \frac{\eta}{p^+} - \frac{i\pi}{2}
  - i  \pi \ C_\infty + i \pi C_\infty \)
    \]
\nonumber \\
& = &
  - \frac{\a_s}{\pi}\ C_{\rm F}\   \frac{1}{\e}
    \[1 - \frac{ \gamma^+ \hat p}{2 p^+}
    \( 1  + \ln \frac{\eta}{p^+} - \frac{i\pi}{2} \) \]
    \ .
\label{eq:s_tot}
\end{eqnarray}

 \begin{figure}[t]
\centering
\includegraphics[scale=0.5,angle=90]{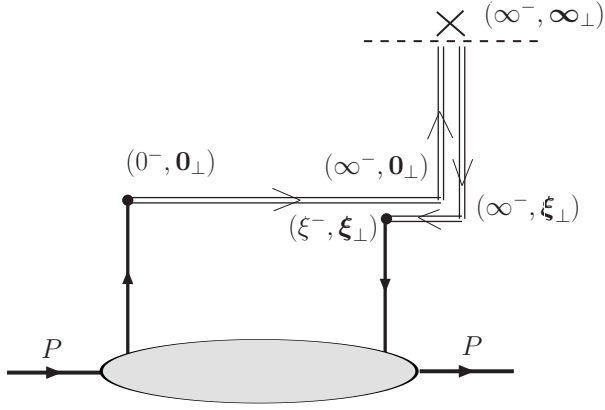}~~
\caption{Graphical representation of a generic TMD PDF (shaded oval)
         in coordinate space.
         The double lines denote the lightlike and transverse gauge
         links, connecting the quark field points
         $(0^{-},{\bf 0}_{\perp})$ and
         $(\xi^{-},\mbox{\boldmath$\xi_\perp$})$,
         by a composite contour through light-cone infinity.
         The latter is marked by the typical symbol for the ground in
         an electrical circuit.
         The contour obstruction at infinite transverse and lightlike
         distance $(\infty^-, \mbox{\boldmath$\infty_\perp$})$
         is symbolized by a cross, whereas the broken line
         indicates that this obstruction is ``hidden''.
\label{fig:generic-TMD-PDF}}
\end{figure}

Next, taking into account that
$$
  \frac{ \gamma^+ \hat p \gamma^+}{2 p^+} = \gamma^+
$$
and recalling that the mirror (which we termed before ``right'')
counterparts of the evaluated diagrams yield the complex-conjugated
contributions, one can conclude that the imaginary terms above
mutually cancel, so that the UV-divergent part of diagrams (a) and (d)
contains only contributions due to the $p^+$-dependent term, notably,
\begin{equation}
  \Sigma_{\rm UV}^{\rm (a+d)}(\alpha_s, \epsilon)
=
   2\frac{\alpha_s}{\pi}C_{\rm F} \left[ \frac{1}{\epsilon}
   \left( \frac{3}{4}
  + \ln \frac{\eta}{p^+} \right) - \gamma_E + \ln 4\pi \right]\, ,
\label{eq:gamma_1}
\end{equation}
plus those terms originating from the standard \MS renormalization.
Hence, there is an extra anomalous dimension associated with the
$p^+$-dependent term which at the one-loop level, considered here, is
given by
\begin{equation}
  \gamma_{\rm 1-loop}^{\rm LC}
=
  \frac{\alpha_s}{\pi}C_{\rm F}\Bigg( \frac{3}{4}
  + \ln \frac{\eta}{p^+} \Bigg)
=
  \gamma_{\rm smooth} - \delta \gamma \ .
\label{eq:gamma_2}
\end{equation}
The difference $\delta \gamma$ between $\gamma_{\rm smooth}$ and
$\gamma_{\rm LC}$ is exactly that term induced by the additional
divergence which ultimately has to be compensated by a suitable
redefinition of the TMD PDF, if we want to reproduce the same
anomalous dimension as in a covariant gauge.

Here, some comments are in order.
It was shown (see, e.g., \cite{BR05}) that the Mandelstam-Leibbrand
(ML) prescription \cite{Man82, Lei83, Lei87}
\begin{equation}
  \frac{1}{[q^+]_{\rm ML}}
  =
  \frac{1}{q^+ + i 0\ q^-}
  =
  \frac{q^-}{q^+q^- + i0} \
\end{equation}
yields a $p^+$-independent anomalous dimension of the quark fields,
i.e.,
\begin{equation}
   \gamma_{\rm ML}^{\rm LC}
=
   \frac{3}{4} \frac{\alpha_s}{\pi}C_{\rm F} + O(\alpha_s^2) \ .
\end{equation}
Moreover, the ML-prescription, in contrast to the PV/Adv/Ret ones
(cf.\ Eq.\ (\ref{eq:c_inf})), entails additional poles in the complex
$q^0$-plane which allow for a Wick rotation and, therefore, it does
not break the standard power counting rules.
However, on the one hand, it is not clear how the ML-prescription can
be related to any boundary conditions on the gauge field at light-cone
infinity, thus making the popular initial/final state interactions
interpretation questionable.
On the other hand, the ML-regularization appears to be not sufficient
for the calculation of the transverse gauge field at light-cone infinity
(in the form of an expression analogous to, say, Eq.\
(\ref{eq:pole-prescription})).
The latter issue is potentially crucial for reproducing the results
obtained in covariant gauges, while within the PV/Adv/Ret methods,
the similarity between the light-cone and covariant gauges can be
explicitly established.
These issues will be further investigated and quantified in a 
separate work.

To continue, recall that in a covariant gauge the gluon field 
vanishes at infinity and, hence, the only anomalous dimensions ensuing 
from the gauge link stem from its endpoints that are joined by a smooth 
direct contour.\footnote{It is worth reiterating that all smooth gauge
contours yield the same anomalous dimensions, $\gamma_{\rm smooth}$,
as the straight line between the endpoints, because only the latter
are relevant.}
Actually, $p^+ = (p \cdot n^-) \sim \cosh \chi$ defines an angle $\chi$
between the direction of the quark momentum $p_\mu$ and the lightlike
vector $n^-$.
Then, in the large $\chi$ limit, one has
$\ln p^+ \to \chi , \ \chi \to \infty$.
Thus, we come to recognize that the ``defect'' of the anomalous
dimension, $\delta \gamma$, can be identified with the well-known cusp
anomalous dimension \cite{KR87}
\begin{equation}
\begin{split}
   & \gamma_{\rm cusp} (\alpha_s, \chi)
= \frac{\alpha_s}{\pi}C_{\rm F} \ (\chi \coth \chi - 1 ) \ , \\
& \frac{d}{d \ln p^+} \ \delta \gamma
= \lim_{\chi \to \infty}
  \frac{d}{d \chi} \gamma_{\rm cusp} (\alpha_s, \chi)
= \frac{\alpha_s}{\pi}C_{\rm F} \ .
\end{split}
\end{equation}

\begin{figure}[t]
\centerline{%
\includegraphics[width=0.28\textwidth]{%
                 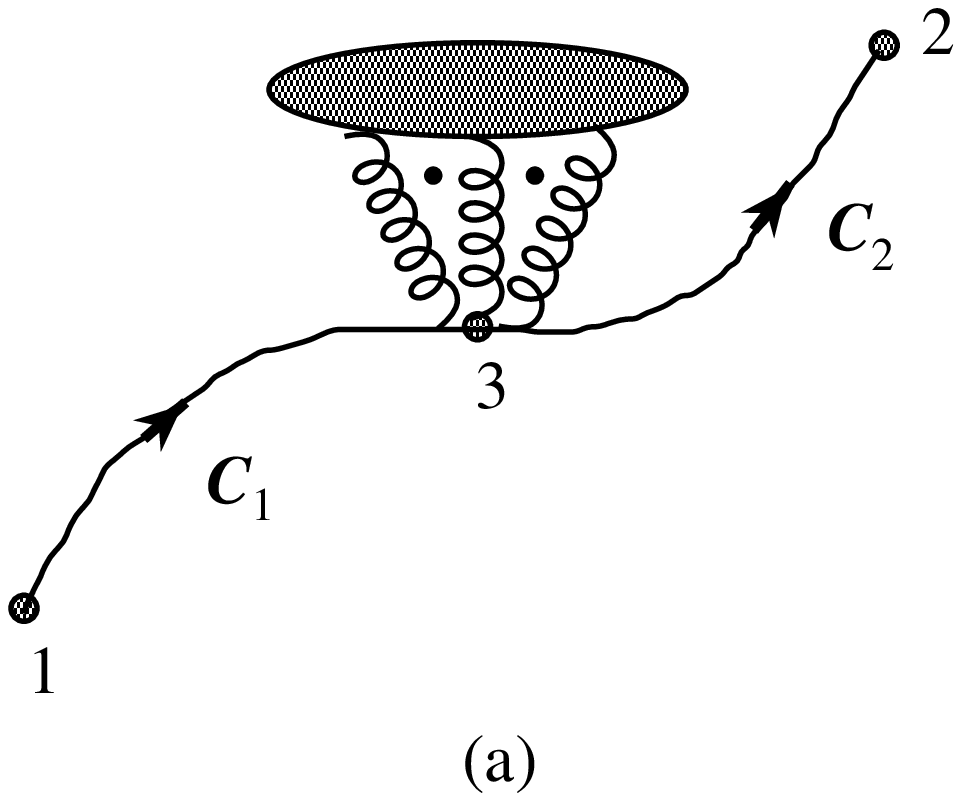}
~~~~~~~~~~~~~~~~~~~~~~%
\includegraphics[width=0.26\textwidth]{%
                 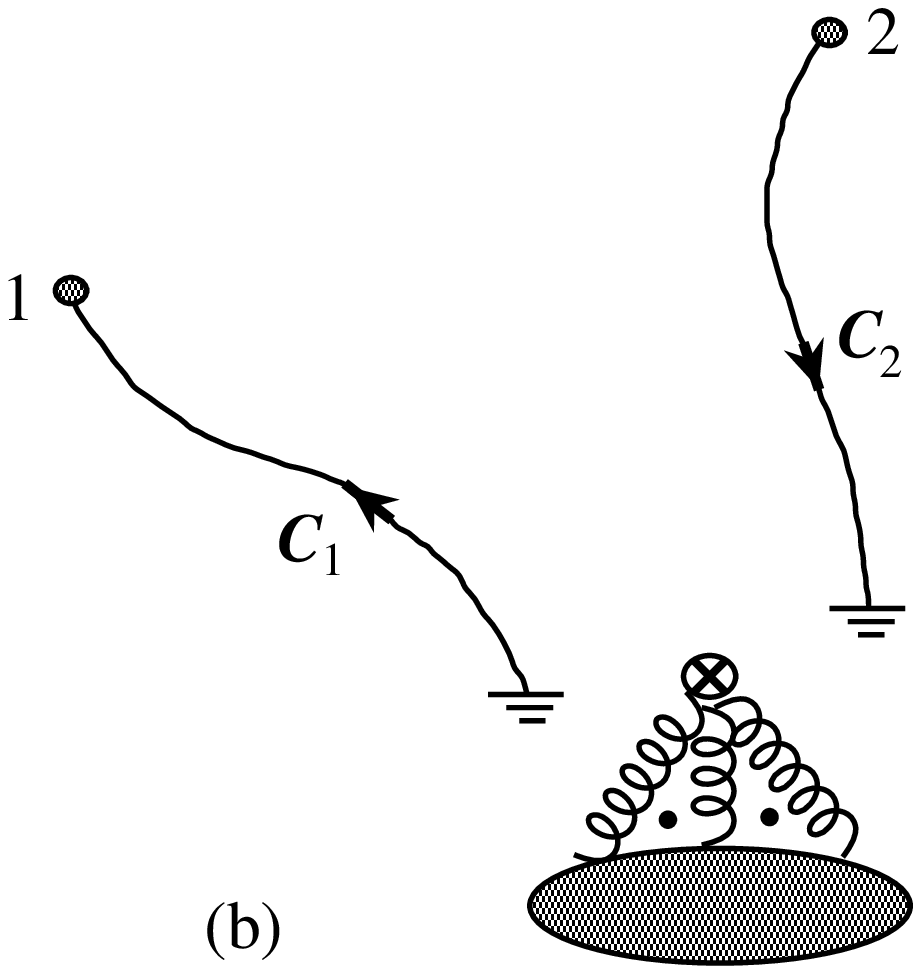}}
\caption{Renormalization effect on the junction point due to gluon
         corrections (illustrated by a shaded oval with gluon lines
         attached to it) for (a) two smoothly joined gauge contours
         $\mathcal{C}_1$ and $\mathcal{C}_2$ at point 3 and (b) the
         same for two contours joined by a cusp (indicated by the
         symbol $\otimes$) at infinite transverse distance (marked
         by the earth symbol) off the light cone.
         All contours shown are assumed to be arbitrary non-lightlike
         paths in Minkowski space.
\label{fig:cont-fact}}
\end{figure}

This is an important observation that deserves to be discussed in
some detail.

As we mentioned earlier, splitting the gauge contour for non purely
lightlike contours through the light-cone infinity, is not equivalent
to the situation with a direct contour between the two field points.
To understand the deeper reason for this difference, we have to study
again the algebraic identity (\ref{eq:link-ident}) for decomposing
(factorizing) gauge contours (links).
The crucial question here is whether the defect of the anomalous
dimension, we calculated, is compatible with this identity when the
junction point is assumed to be at infinite distance in the transverse
configuration space.
To answer this question, consult Fig.\ \ref{fig:cont-fact}.
Panel (a) of this figure shows the renormalization effect (illustrated
by a shaded oval with gluon lines attached to it) on the junction point
in the algebraic identity (\ref{eq:link-ident}).
The contour $\mathcal{C}_1\cup \mathcal{C}_2$ is smooth and
non-self-intersecting owing to the assumption that $\mathcal{C}_1$ and
$\mathcal{C}_2$ are smoothly connected at 3.
Then, both contours the direct one, $\mathcal{C}$, and the decomposed
one, $\mathcal{C}_1\cup \mathcal{C}_2$, between the endpoints 1 and 2,
cannot be distinguished from each other by switching on gluon quantum
corrections.
In particular, no anomalous dimension emerges from the junction point
3 and thus (symbolically)
\begin{equation}
  \gamma_{\mathcal{C}}
=
  \gamma_{\mathcal{C}_1\cup\, \mathcal{C}_2} \ .
\label{eq:an-direct}
\end{equation}
Now we may ask what changes are induced, if we allow the junction
point 3 to be shifted to infinity in the transverse direction off the
light cone.
The graphics at right of Fig.\ \ref{fig:cont-fact} helps the eye catch
the key features of the situation involving two non-lightlike contours
$\mathcal{C}_1$ and $\mathcal{C}_2$.
It turns out that the naive assumption that
\begin{equation}
  \gamma_{\mathcal{C}}
=
  \gamma_{\mathcal{C}_1^{\infty}\cup\, \mathcal{C}_2^{\infty}}
\label{eq:an-split-inc}
\end{equation}
is incorrect for contours containing transverse segments.
Instead, we found that in this case
\begin{equation}
   \gamma_{\mathcal{C}}
=
   \gamma_{\mathcal{C}_1^{\infty}\cup\, \mathcal{C}_2^{\infty}}
  +\gamma_{\rm cusp} \ .
\label{eq:an-direct-corr}
\end{equation}
Consequently, the validity of the algebraic identity
((\ref{eq:link-ident})) is not conserved and we have to replace it
by the generalized gauge-link factorization rule
\begin{equation}
  [2,1|\mathcal{C}]
=
  [2,\infty|\mathcal{C}_{2}^{\infty}]^\dag
  [\infty,1|\mathcal{C}_1^{\infty}]
  {\rm e}^{i \Phi_{\rm cusp}} \ ,
\label{eq:gauge-link-fact}
\end{equation}
which is valid for arbitrary paths in Minkowski space.
In this expression, $\Phi_{\rm cusp}$ takes care of the effect induced
by the cusp-like junction point.
We will consider an explicit example of such a phase in the next
subsection, where we show how to compensate it by an eikonal factor
in the definition of the TMD PDF.
One may associate this phase with final (or initial) state
interactions, as proposed by Ji and Yuan in \cite{JY02}, and also
by Belitsky, Ji, and Yuan in \cite{BJY02}.
However, these authors (and also others) did not recognize that the
junction point in the split contour
(taking a detour to light-cone infinity)
is no more a simple point, but becomes a cusp obstruction
$\sim \ln p^+$ that entails an anomalous dimension as the result of a
non-trivial renormalization effect owing to gluon radiative
corrections.

These arguments make it clear that the naive decomposition of
gauge contours that stretch out to light-cone infinity along the
transverse direction is erroneous, simply because the basic
algebraic identity (\ref{eq:link-ident}), which is tacitly assumed,
is inapplicable to such contours and has to be replaced by Eq.\
(\ref{eq:gauge-link-fact}).
It almost goes without saying that the modified factorization rule,
expressed through Eqs.\ (\ref{eq:an-direct-corr}) and
({\ref{eq:gauge-link-fact}), is valid when one is composing
\emph{non-smoothly} any gauge contours with a cusp obstruction at the
junction point.\footnote{A similar factorization rule holds for
contours joined through a self-crossing point.}
What marks out a cusped contour from all the others, however, is that
it gives rise to an anomalous dimension proportional to $\ln p^+$,
i.e., to a jump in the four-velocity.
This is a salient ingredient in describing correctly a DIS process in
spacetime, because if the two quarks (the struck one and a spectator)
are separated also in the transverse coordinate space, the gluons
emitted mismatch in rapidity and, hence, the contour liaising them
has to have a sharp bend and cannot be the direct one.

\subsection{Compensating the defect of the anomalous dimension by a
            soft counter term}
\label{subsec:soft-anom-dim}

The defect of the anomalous dimension owing to the gauge-contour cusp
at light-cone infinity represents a distortion of the gauge-invariant
formulation of the TMD PDF in the light-cone gauge.
To restore its consistency, we have to dispense with the
anomalous-dimension artefact of the cusp.
This can be achieved by supplying the original definition of
$f_{q/q}(x, \mbox{\boldmath$k_\perp$})$
by a soft counter term in the sense of Collins and Hautmann
\cite{CH99,CH00,Hau07,HJ07}:
\begin{equation}
  R
\equiv
 \Phi (p^+, n^- | 0) \Phi^\dagger (p^+, n^- | \xi) \ ,
\label{eq:soft_factor_1}
\end{equation}
where the eikonal factors are given by
\begin{eqnarray}
\Phi (p^+, n^- | 0 )
 & = &
  \left\langle 0
  \left| {\cal P}
  \exp\Big[ig \int_{\mathcal{C}_{\rm cusp}}\! d\zeta^\mu
           \ t^a A^a_\mu (\zeta)
      \Big]
  \right|0
  \right\rangle \
\ , \\
  \Phi^\dagger (p^+, n^- | \xi )
 & = &
  \left\langle 0
  \left| {\cal P}
  \exp\Big[- ig \int_{\mathcal{C}_{\rm cusp}}\! d\zeta^\mu
           \ t^a A^a_\mu (\xi + \zeta)
      \Big]
  \right|0
  \right\rangle \ ,
\label{eq:soft_definition}
\end{eqnarray}
and evaluate $R$ along the non-smooth (non-lightlike) integration
contour
$\mathcal{C}_{\rm cusp}$,
defined by
\begin{equation}
 \mathcal{C}_{\rm cusp}:\zeta_\mu
=
  \left\{
         [p_\mu^{+}s, - \infty < s < 0]
         \cup [n_\mu^-  s^{\prime},
         0 < s^{\prime} < \infty] \cup
         [ \mbox{\boldmath$l_\perp$} \tau , 0 < \tau < \infty ]
  \right\} \ ,
\label{eq:gpm}
\end{equation}
$\rule{0in}{2.6ex}$
with $n_\mu^-$ being the minus light-cone vector, as illustrated in
Fig.\ \ref{fig:contour}.

Contour (\ref{eq:gpm}) is obviously cusped: at the origin, the
four-velocity $p_\mu^{+}$, which is parallel to the plus
light-cone ray, is replaced---non-smoothly---by the four-velocity
$n_\mu^-$, which is parallel to the minus light-cone ray.
This jump in the four-velocity becomes visible in the standard
leading-order term
\begin{equation}
  g^2 \int_0^\infty \! ds \int_0^\infty \!ds^{\prime}
  \frac{(v_1 \cdot v_2)}{(v_1 s - v_2 s^{\prime})^2}
=
  g^2 \int_0^\infty \! ds \int_0^\infty \!ds^{\prime}
  \frac{(v_1 \cdot v_2)}{v_1^2 s + v_2^2 s^{\prime}
        - 2 (v_1 \cdot v_2) s s^{\prime}} \ ,
\label{eq:leading-order-term}
\end{equation}
in which $v_1 = p^+ , \ v_2 = n^-$, and the change of the four-velocity
at the origin produces an angle-dependence via $(v_1 \cdot v_2) = p^+$.
This means that exactly at this point the contour has a cusp that is
characterized by the angle
$\chi \sim \ln p^+ = \ln (p \cdot n^-)$,
and, therefore, the corresponding eikonal factor
(\ref{eq:soft_definition}) gives rise to a cusp anomalous
dimension.
Obviously, this is exactly what we need in order to compensate the
extra term in the anomalous dimension found in the preceding
subsection.

\begin{figure}[t]
\centering
\includegraphics[scale=0.6,angle=0]{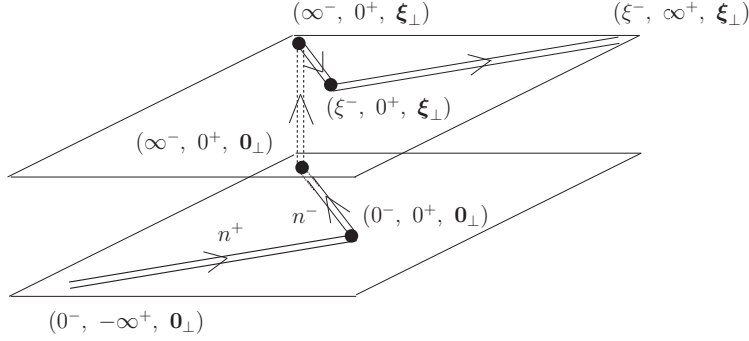}~~
\caption{The integration contour associated with the additional soft
         counter term.
\label{fig:contour}}
\end{figure}

Next, we show that the one-loop gluon virtual corrections, contributing
to the UV divergences of $R$ and displayed in Fig.\
\ref{fig:soft_gluon}, yield an anomalous dimension that neutralizes the
cusp artefact $\delta\gamma$.
Note that in the light-cone gauge $A^+ = (n^- \cdot A) = 0$ only the
first lightlike ray $- \infty < s < 0$ and also the transverse segment
contribute, since the other eikonal line along the minus lightlike ray
depends on the longitudinal component of the gauge field and vanishes
due to the gauge condition.

Calculate first the diagram $(a)$ in Fig.\ \ref{fig:soft_gluon}.
In leading order, the first nontrivial term in Eq.\
(\ref{eq:soft_definition}) reads
\begin{equation}
\begin{split}
  \F^{\rm (1-loop)}_a(u, \eta)
= {} &
  i g^2 \m^{2\e} C_{\rm F}\ u_\m u_\n
  \int_0^\infty d\s \int_0^\s d\t \int\! \frac{d^\w q}{(2\pi)^\w}
  \frac{\ex^{-i q \cdot u (\s-\t)}}{q^2 - \lambda^2}
  \(g^{\m\n} - \frac{q^\m n^{-\n} + q^\n n^{-\m}}{[q^+]} \)
\\
= {} &
   i g^2 \m^{2\e} C_{\rm F}
   \int\! \frac{d^\w q}{(2\pi)^\w} \frac{1}{q^2 - \lambda^2}
   \left[ - \frac{{u^2}}{(q \cdot u - i0)^2}
  +\frac{2 u^+} {(q \cdot u - i0)[q^+]}
   \right] \ .
   \end{split}
\label{eq:soft-factor1}
\end{equation}
The first term in the square bracket vanishes since $u_\m$ is chosen
to point along the $p^+$-direction, i.e.,
$u_\m = (p^+, 0^-, \vecc 0_\perp)\ , \  u^2 = 0 $,
and by recalling that in dimensional regularization
$u^2/(u^{2-\e}) = 0$.
Notice that in a covariant gauge this diagram would be tantamount to
the self-energy contribution of the struck quark.
However, in the light-cone gauge, we are employing, a second term
in the parenthesis in the first line of Eq.\ (\ref{eq:soft-factor1})
appears which stems from the gluon propagator in that gauge.
This term, being not lightlike, also entails a contribution to the
cusp-dependent part, as we will now show.
Indeed, one has
\begin{equation}
  \F^{\rm (1-loop)}_a(u, \eta)
=
  i g^2 \m^{2\e} C_{\rm F}2 p^+ \ \int\! \frac{d^\w q}{(2\pi)^\w}
  \frac{1}{(q^2 - \lambda^2) (q \cdot u - i0) [q^+]} \ ,
\label{eq:f1}
\end{equation}
an expression which would correspond to a vertex-like contribution
of the pure gauge link in a covariant gauge---as one may appreciate
from Eq.\ (\ref{eq:leading-order-term}).

The pole-prescription dependent integral can be evaluated in analogy
to our previous calculations in the preceding subsection, so that
\begin{equation}
  \F^{\rm (1-loop)}_a(u, \eta)
=
  - g^2 C_{\rm F}2  \(\frac{4\pi \m^2}{\lambda^2}\)^\e
  \frac{\G(\e)}{(4\pi)^2}
  \int_0^1 dx \frac{1}{x^2\[-\frac{\bar x}{2x}\]} \ ,
\label{eq:soft_int}
\end{equation}
where the bracketed term in the denominator is to be evaluated with
the aid of Eq.\ (\ref{eq:pole-prescr}).
The last step in obtaining an explicit expression for
$\F^{\rm (1-loop)}_a(u, \eta)$
is to carry out the line integral over $x$ (which enters because of
the appearance of the gauge link in the lightlike direction).
To do so, we have to take care of the additional (logarithmic)
singularity $\sim \ln \tau$, which cannot be regularized by the
parameter $\eta$, where $\tau$ is an extra regulator.
The origin of this singularity is related to the vector $u_\m$ which
defines the lightlike direction.
There are, of course, several possibilities how to regularize this
type of integral.
For instance, one can get a regular expression in $\tau$---after the
integration over $d\eta$---as discussed in Refs.\
(\cite{KaKo92}, \cite{BKKN93}) having recourse to the fact that the
derivative
$ \partial \F^{\rm (1-loop)}_a(u, \eta)/\partial \eta $
is $\tau$-independent.
However, for technical convenience, we apply here a different
regularization technique by making use of an auxiliary regulator
$\tau$ and absorb the light-cone singularity $\ln \tau$ inside a
redefined parameter $\tilde \eta = 2 \tau \eta$, the latter being
contained inside the pole-prescription contribution---cf.\ Eq.\
(\ref{eq:pole-prescr}).
This is possible, given that $\tau$ does not depend on the scale
parameter $\eta$ and hence does not contribute to the evolution
of the considered quantity (see Sec.\ \ref{sec:evolution}).

\begin{figure}[t]
\centering
\includegraphics[scale=0.60,angle=90]{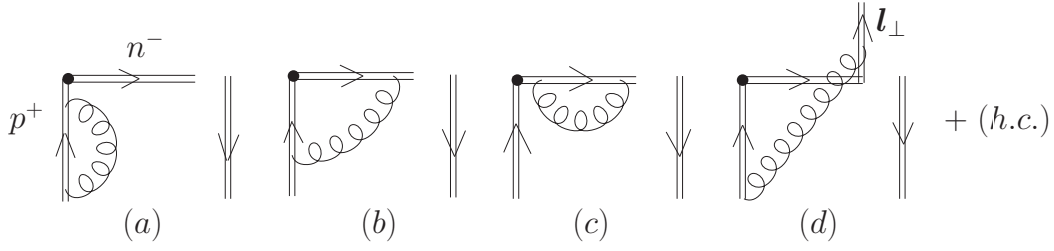}~~
\caption{Virtual gluon contributions to the UV-divergences of the soft
         counter term, given by Eq.\ (\ref{eq:soft_factor_1}).
         The designations are as in Fig.\ \ref{fig:se_gluon}.
\label{fig:soft_gluon}}
\end{figure}

Performing all these operations and taking into account that
$u^+ = p^+$, one gets for the UV part of diagram (a) in Fig.\
\ref{fig:soft_gluon}
\begin{equation}
  \F^{\rm (1-loop)}_{\rm UV}(\eta)
=
  - \frac{\a_s}{\pi} C_{\rm F} \frac{1}{\e}
  \( \ln \frac{\eta}{p^+} - i\frac{\pi}{2} - i \pi C_\infty \) \ .
\label{eq:soft_a}
\end{equation}
Evidently, this expression yields a cusp-dependent contribution,
$\sim \ln p^+$, as we already mentioned, which would be completely
absent in a covariant gauge, thus, underlying their mutual difference.
Besides, there is a dependence on the choice of the pole prescription
(via the numerical parameter $C_\infty$).
In order to cancel this latter dependence, one needs to take into
account the contribution of diagram (d) in Fig.\
\ref{fig:soft_gluon}; viz.,
\begin{eqnarray}
  \F^{\rm (d)}
 = &&\!\!\!
  - g^2C_{\rm F} \m^{2\e} p^+
  \int\! \frac{d^\w q}{(2\pi)^\w}
  \int\! \frac{dq^{\prime +}}{2\pi} \ex^{- i q^{\prime +} \infty^-}\!\!
  \int\! \frac{d^2 q^{\prime}_\perp}{(2\pi)^2}
  \frac{(\vecc l_\perp \cdot \vecc q_\perp)}{(q^2 - \lambda^2 ) [q^+]}
\nonumber \\
&& \!\times
  (-i) (2\pi)^4\d^{(4)} (q+q^{\prime})
  \frac{1}{q \cdot u + i0 }
  \frac{1}{\vecc q_{\perp}^{\prime} \cdot \vecc l_\perp + i0} \ .
\label{eq:dia-d-1}
\end{eqnarray}
Using Eqs.\ (\ref{eq:trans_1})--(\ref{eq:trans_4}), we find
\begin{equation}
\begin{split}
   \F^{\rm (d)}
=  i g^2 C_{\rm F} \m^{2\e} \pi C_\infty \int\! \frac{d^{\w -2}
   q_\perp}{(2\pi)^{\w-2}} \frac{1}{\vecc q_\perp^2 + \l ^2}
=
   - \a_s C_{\rm F} i \pi C_\infty \G(\e)
   \(- 4\pi \frac{\m^2}{\l^2}\)^\e
\end{split}
\end{equation}
and extracting the UV-pole and adding it to Eq.\ (\ref{eq:soft_a}) we
finally arrive at
\begin{equation}
  \F^{\rm (a + d)}_{\rm UV}(\eta)
=
  - \frac{\a_s}{\pi} C_{\rm F} \frac{1}{\e}
  \( \ln \frac{\eta}{p^+} - i\frac{\pi}{2}
    - i \pi C_\infty + i \pi C_\infty
  \)
=
   - \frac{\a_s}{\pi}C_{\rm F}\frac{1}{\e}
  \( \ln \frac{\eta}{p^+} - i\frac{\pi}{2}
  \) \ .
\label{eq:soft_ad}
\end{equation}
This result exhibits the independence on the pole prescription
of the soft factor $R$, in close analogy to Eq.\ (\ref{eq:gamma_2}).
Taking into account the corresponding ``mirror'' diagram
(which doubles the real part and cancels the imaginary one),
we obtain the total UV-divergent part of the soft factor in the
one-loop order:
\begin{equation}
  \F^{\rm (1-loop)}_{\rm UV}(\eta)
  =
  - \frac{\a_s}{\pi}C_{\rm F}  \frac{2}{\e}
  \ln \frac{\eta}{p^+} \ ,
  \label{eq:phase_1loop}
\end{equation}
making it apparent that there is no dependence on the pole
prescription, as now all $C_\infty$-dependent terms are absent.

To conclude, we have shown at the one-loop level that the soft
counter term (soft eikonal factor) has the following two important
properties:\\
(i) it gives rise to the same cusp anomalous dimension as
$f_{q/q}(x, \mbox{\boldmath$k_\perp$})$,
but with an opposite sign, and \\
(ii) it bears no dependence on the choice of the pole prescription to
go around the light-cone singularity in the light-cone gauge
(with corresponding terms cancelling among themselves).

Therefore, it is reasonable to redefine the conventional TMD PDF by
including into its definition the soft counter term ab initio.
This provides
\begin{eqnarray}
  f_{q/q}^{\rm mod}\left(x, \mbox{\boldmath$k_\perp$};\mu, \eta\right)
&& \!\!\! =
  \frac{1}{2}
  \int \frac{d\xi^- d^2\mbox{\boldmath$\xi_\perp$}}{2\pi (2\pi)^2}
  {\rm e}^{-ik^+\xi^- +i\bit{\scriptstyle k_\perp}
  \cdot\bit{\scriptstyle\xi_\perp}}
  \left\langle
              q(p) |\bar \psi (\xi^-, \mbox{\boldmath$\xi_\perp$})
              [\xi^-, \mbox{\boldmath$\xi_\perp$};
   \infty^-, \mbox{\boldmath$\xi_\perp$}]^\dagger
\right. \nonumber \\
&& \left. \times
   [\infty^-, \mbox{\boldmath$\xi_\perp$};
   \infty^-, \mbox{\boldmath$\infty_\perp$}]^\dagger
   \gamma^+[\infty^-, \mbox{\boldmath$\infty_\perp$};
   \infty^-, \mbox{\boldmath$0_\perp$}]
   [\infty^-, \mbox{\boldmath$0_\perp$}; 0^-,\mbox{\boldmath$0_\perp$}]
\right. \nonumber \\
&& \left. \times
   \psi (0^-,\mbox{\boldmath$0_\perp$}) |q(p)
   \right\rangle
   \Big[ \Phi(p^+, n^- | 0^-, \mbox{\boldmath$0_\perp$})
   \Phi^\dagger (p^+, n^- | \xi^-, \mbox{\boldmath$\xi_\perp$})
   \Big] \, ,
\label{eq:tmd_re-definition}
\end{eqnarray}
which represents one of the main results of our investigation here
and in \cite{CS07}.

Let us finish this section by giving a physical interpretation to
the soft counter term, using Mandelstam's formalism
\cite{Man62,Man68YM}.
To this end, we utilize the exponentiation theorem for non-Abelian
path-ordered exponentials \cite{KR87} and recast the exponential
operator (\ref{eq:soft_definition}) in the form
\begin{equation}
  \Phi (u, n^-)
=
  \exp \left[\sum_{n=1}^{\infty} \alpha_s^n \Phi_n (u, n^-)
       \right] \ ,
\end{equation}
where the functions $\Phi_n$ have, in general, a complicated
structure that is, however, irrelevant for our purposes here.
The leading term in this series, $\Phi_1$, is just a non-Abelian
generalization of the Abelian expression
\begin{equation}
  \Phi_1 (u, n^-)
= - 4 \pi C_{\rm F}
  \int_{\mathcal{C}_{\rm cusp}}\! dx_\mu dy_\nu
  \theta (x-y) D^{\mu\nu} (x-y) \ .
\label{eq:phase_1}
\end{equation}
Then, by virtue of the current
\begin{equation}
  j_\nu^b (z)
=
  t^b v_\nu \int_{\mathcal{C}_{\rm cusp}}
  d\tau \delta^{(4)}(z - v\tau) \ ,
\end{equation}
evaluated along the contour
$\mathcal{C}_{\rm cusp}$ (cf.\ Eq.\ (\ref{eq:gpm})) and where the
velocity $v_\nu$ equals either $u_\nu$, $n^-$, or
$\mbox{\boldmath$l_\perp$}$
(depending on the segment of the contour along which the integration
is performed), one can rewrite (\ref{eq:phase_1}) as follows
\begin{equation}
  \Phi_{1}(u, n^-)
=
  \ - t^a 4\pi
  \int_{\mathcal{C}_{\rm cusp}}\! dx_\mu  \int d^4 z
  \delta^{ab} D^{\mu\nu} (x-z) j_\nu^b (z) \ .
\label{eq:coul}
\end{equation}
This expression looks formally very similar to the ``intrinsic''
Coulomb phase found by Jakob and Stefanis (JS) \cite{JS91} in QED
for Mandelstam charged fields involving a gauge contour which is a
timelike straight line.
The name ``intrinsic'' derives from the fact that this phase is
different from zero even in the absence of external charge
distributions.
Its origin was ascribed by JS to the long-range interaction of
the charged particle with its oppositely charged counterpart that
was removed ``behind the moon'' after their primordial
separation.\footnote{The existence of a balancing charge ``behind
the moon'' was postulated before by several authors---see \cite{JS91}
for related references---in an attempt to restore the Lorentz
covariance of the charged sector of QED.}
This phase is acquired during the parallel transport of the charged
field along a timelike straight line from infinity to the point of
interaction with the photon field and is absent in the local approach,
i.e., for local charged fields joined by a connector.
It is different from zero only for Mandelstam fields with their own
gauge contour attached to them and keeps track of its full history
since its primordial creation.
Keep in mind that the connector is introduced ad hoc in order to
restore gauge invariance and is not part of the QCD Lagrangian.
In contrast, when one associates a distinct contour with each quark
field, one, actually, implies that these Mandelstam field variables
should also enter the QCD Lagrangian (see \cite{JS91} for more
details).
However, a consistent formulation of such a theory for QCD is still
lacking and not without complications of its own.

The analogy to our case is the following.
First, formally adopting a direct contour for the gauge-invariant
formulation of the TMD PDF in the light-cone gauge
(Figure \ref{fig:con-unsplit_split}(a) shows an example of the
contributing diagrams), the connector gauge link does not contribute
any anomalous dimension---except at the endpoints; this anomalous
dimension being, however, irrelevant for the issue at stake.
Hence, there is no intrinsic Coulomb phase in that case.
Second, splitting the contour and associating each branch to a
quark field, transforms it into a Mandelstam field and, as a result,
adding together all gluon radiative corrections at the one-loop order,
a $p^+$-dependent term survives that gives rise to an additional
anomalous dimension.
We have shown that this extra anomalous dimension can be viewed as
originating from a contour with a discontinuity in the four-velocity
$\dot x(\sigma)$ at light-cone infinity---a cusp obstruction.

Classically, it is irrelevant how the two distinct contours
$\mathcal{C}_1$ and $\mathcal{C}_2$ in Fig.\ \ref{fig:cont-fact}
are joined, i.e., smoothly or by a sharp bend.
But switching on gluon quantum corrections, the renormalization
effect on the junction point reveals that the contours are not
smoothly connected, but go instead through a cusp.
Here, we have a second analogy to the QED case discussed above.
Similarly to the ``particle behind the moon'', this cusp-like junction
point is ``hidden'' and manifests itself only through the
path-dependent phase (\ref{eq:coul}).
Note in the same context that integrating over the transverse momentum
(see next section), the $p^+$-dependent terms, resulting from virtual
gluon corrections, cancel against their counterparts from real-gluon
corrections, so that this cusp-induced phase disappears [see for
illustration Fig.\ \ref{fig:con-unsplit_split}(b)].

\begin{figure}
\centering
\includegraphics[scale=0.60,angle=90]{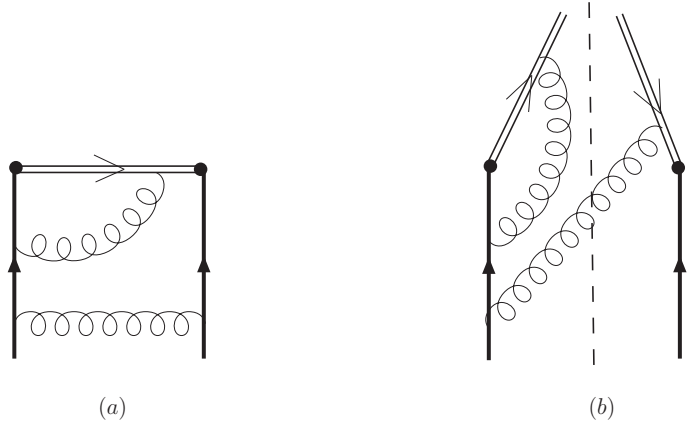}~~
\caption{(a) Schematic illustration of the direct gauge contour with
         virtual gluon-line insertions (denoted by curly lines).
         (b) Similar illustration for real gluon exchanges over the
         cut, the latter indicated by a dashed line.
         In both panels double lines
         represent the gauge links.
\label{fig:con-unsplit_split}}
\end{figure}

In our previous paper \cite{CS07}, we concentrated on the anomalous
dimension of the TMD PDF, and, therefore, only the UV-divergent parts
were studied.
In the present work, however, we take also into account the UV-finite
parts and, consequently, the dependence on the transverse momentum
appears explicitly, as we discussed above.

\section{Real-gluon contributions and evolution equations}
\label{sec:evolution}

In this section, we concentrate our efforts on two subjects:
(i) First we discuss in some detail the evolution behavior of the
TMD PDF and establish the connection between our approach and that
of Collins and Soper \cite{CS81}.
(ii) Second, we prove that the integrated PDF, obtained from our
modified definition, coincides with the standard one with no any
artefact of the cusped contour used in the TMD PDF left over.

(i) \emph{Evolution behavior.}
The modified TMD PDF (\ref{eq:tmd_re-definition}) depends on two
arbitrary mass-scale parameters: the UV scale $\mu$ and the extra
regulator $\eta$.
The $\mu$-dependence is described by the standard renormalization-group
evolution equation (see below) and is controlled by the
UV-anomalous dimension, which arises as the sum of the
anomalous dimensions of all the ingredients of the TMD PDF
(\ref{eq:tmd_re-definition}):
\begin{eqnarray}
  \gamma_{f_{q/q}}
& = &
   \gamma_{2q} + \sum_{i=1}^4
   \gamma_{\rm gauge\ link}^i + \gamma_{\rm R}  \nonumber \\
& = &
   \frac{3}{4} \frac{\a_s}{\pi} C_{\rm F} + O(\alpha_s^2) \ .
\label{eq:anom-dim}
\end{eqnarray}
The anomalous dimensions associated with the quark fields and the
soft counter term $R$ are marked by self-explaining labels.
We have used for convenience a short-hand notation to denote the
anomalous dimension of each gauge link on the right-hand side of Eq.\
(\ref{eq:tmd_re-definition}) by a number in the order the gauge link
appears from the left to the right.

Before we proceed, a couple of important remarks are here in order.
One realizes that the anomalous dimension of $f_{q/q}$ coincides with
the anomalous dimension of the conventional quark propagator in
the light-cone gauge, but with the opposite sign due to the different
Dirac structure.
Up to the sign, this result also coincides with the anomalous
dimension of the gauge-invariant quark propagator in a covariant gauge
\cite{Ste83}.
The anomalous dimension of $R$, $\gamma_R$, cancels precisely those
contributions in the sum above which contain the $p^+$-dependent terms.

On the other hand, the dependence on $\eta$ is more complicated and
is described by an integral kernel to be determined below.
In order to derive the corresponding evolution equation, one needs to
calculate the real-gluon contributions, depicted in
Fig. \ref{fig:real_gluon}.
Here, we present this calculation in the small-$\eta$ limit (that
corresponds to the large-rapidity $\zeta \to \infty$ limit within the
Collins-Soper approach \cite{CS81}).
Let us emphasize that in the case of the integrated PDFs, where the
dependence on the regularization parameter $\eta$ appears at the
intermediate steps of the calculations (in the light-cone gauge),
it cancels out in the final expression.
This will be demonstrated below.
In contrast, in the unintegrated PDFs, this dependence remains and,
thus, it should be treated by means of a corresponding evolution
equation.
\begin{figure}
\centering
\includegraphics[scale=0.60,angle=90]{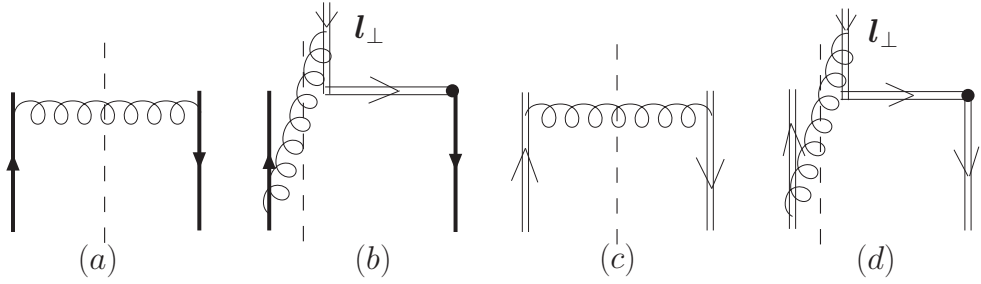}~~
\caption{The leading-order real gluon contributions to the TMD PDF are
         shown.
         The diagrams (b) and (d) with a transverse gauge link do not
         contribute to the TMD PDF in the light-cone gauge.
         The dashed line marks the cut.
\label{fig:real_gluon}}
\end{figure}
Note that the diagrams in Fig. \ref{fig:real_gluon} are UV-finite and
do not contribute to the anomalous dimensions.
However, they do depend on the regularization parameter $\eta$.
The diagram $(a)$ yields
\begin{eqnarray}
  \S^{(a)}_{\rm real}
\!\!\! && =
  - g^2 C_{\rm F} \int\! \frac{d^4 q}{(2\pi)^4}
    \frac{\gamma_\mu (\hat p - \hat q) \gamma^+
    (\hat p - \hat q) \gamma_\nu}{(p-q)^4}
    {\rm Disc} \[D^{\mu\nu} (q)\]
\nonumber \\
&& ~~~\times
    \delta(p^+ - k^+ - q^+)
    \delta^{(2)} (\vecc q_\perp - \vecc k_\perp) \ ,
\label{eq:real_def}
\end{eqnarray}
where the absorptive part of the gluon propagator reads
\begin{equation}
  {\rm Disc} \[D^{\mu\nu} (q)\]
=
    2 \pi \theta(q^+) \delta (q^2 - \lambda^2) \( - g^{\mu\nu}
  + \frac{q^\mu n^{-\nu} + q^\nu n^{-\mu}}{[q^+]_{\rm PV}}\) \ .
\end{equation}
In the last equation we have adopted the PV-prescription,
because the real gluon contributions are prescription-independent
and the diagrams with transverse gauge links do not contribute.
The $\eta$-divergences can be isolated by means of the standard
rules given in \cite{CFP80}:
\begin{equation}
   \frac{1-x}{(1-x)^2 + (\eta /p^+)^2} = -
    \delta(1-x) \ \ln \frac{\eta}{p^+}  + \frac{1}{(1-x)_+} \ .
\end{equation}
After some standard calculations, one gets the ``Feynman''
($\eta$-independent) part
\begin{equation}
  \S^{(a)\rm real}_{\rm Feynman}
=
  \frac{\a_s}{2\pi^2}\, {C_{\rm F}} \frac{|1-x|}{p^+} \
  \frac{ \vecc k_\perp^2 + x \lambda^2
        + x (x-3) p^2}{\left[\vecc k_\perp^2
        + x \lambda^2 - x(1-x) p^2 \right]^2} \ .
\label{eq:real_feyn}
\end{equation}
The $\eta$-dependence appears through the pole-contributions, i.e.,
\begin{equation}
  \S^{(a) \rm real}_{\rm pole}
=
  \frac{\a_s}{\pi^2}\, C_{\rm F}
  \left\{\left[ \frac{x}{(1-x)_+} - \delta(1-x) \ln \frac{\eta}{p^+}
         \right]
   \frac{1}{\vecc k_\perp^2 + x \lambda^2 - x(1-x) p^2 }
  \right\} \ .
   \label{eq:real_a}
\end{equation}
On the other hand, the diagram $(c)$ in Fig. \ref{fig:real_gluon}
yields
\begin{eqnarray}
  \S^{(c)}_{\rm real}
\!\!\! && =
  i g^2\, C_{\rm F}\ u_\m u_\n \ \delta (p^+ - x p^+)
  \int_0^\infty d\s \int_0^\infty d\t \int\! \frac{d^4 q}{(2\pi)^4}
  {\ex^{-i q \cdot u (\s-\t)}}
  \delta^{(2)} (\vecc q_\perp - \vecc k_\perp)
\nonumber \\
&& ~~\times
  2\pi \theta (q^+)
  \delta ({q^2 - \lambda^2})
  \(g^{\m\n} - \frac{q^\m n^{-\n} + q^\n n^{-\m}}{[q^+]_{\rm PV}}
  \)\ .
\end{eqnarray}
In the small-$\eta$ limit, a straightforward calculation gives
\begin{equation}
  \S^{(c)}_{\rm real}
=
  \frac{\a_s}{2\pi^2}\, C_{\rm F}\, \delta (1-x)
  \frac{1}{\vecc k_\perp^2 + \lambda^2 }
  \(1 - \ln \frac{\eta}{p^+}\) \ .
\label{eq:real_c}
\end{equation}
Finally, the (logarithmic) dependence of the modified TMD PDF
on $\eta$ is determined in terms of the equation
\begin{eqnarray}
  \eta \frac{d}{d\eta}
  f_{q/q}^{\rm mod}(x, \mbox{\boldmath$k_\perp$}; \mu, \eta)
\!\!\! && =
  \frac{\alpha_s}{\pi}\, {C_{\rm F}}\, \delta (1-x)
  \[
    \delta^{(2)} (\vecc k_\perp)
    \( \ln \frac{\mu^2}{p^2} - \ln \frac{\mu^2}{\lambda^2} \)
\right. \nonumber \\
&& \left. ~~ -
      \frac{1}{\pi}
      \( \frac{1}{\vecc k_\perp^2 + x \lambda^2 - x(1-x) p^2 }
    + \frac{1}{\vecc k_\perp^2 + \lambda^2 }
      \)
   \] \phi_0 (p) \ .
\label{eq:evo_final}
\end{eqnarray}

We can recast this equation, which governs evolution with respect to
$\eta$, in a form which formally resembles the standard Collins-Soper
evolution equation \cite{CS81,Col89} with respect to $\mu$, namely,
\begin{equation}
  \eta \frac{d}{d\eta}
  f_{q/q}^{\rm mod}(x, \mbox{\boldmath$k_\perp$}; \mu, \eta)
=
  \left[ {\cal K (\mu)} + {\cal G (\mu, \eta) } \right]
  \otimes f_{q/q}^{\rm mod}(x, \mbox{\boldmath$k_\perp$};\mu, \eta) \ .
\label{eq:evo_1}
\end{equation}
The renormalization-group behavior of the functions ${\cal K (\mu)}$
and ${\cal G (\mu, \eta)}$
\cite{GY73} is determined by the universal cusp anomalous dimension
\begin{equation}
  \halb \m \frac{d}{d\m} \ln {\cal K (\mu)}
=
  - \halb \m \frac{d}{d\m} \ln {\cal G (\mu, \eta)}
=
  \gamma_{\rm cusp}
=
  \frac{\a_s}{\pi}\, C_{\rm F} + O(\a_s^2)\ .
\label{eq:ad_evo}
\end{equation}
Extracting explicit expressions for ${\cal K (\mu)}$ and
${\cal G (\mu, \eta)}$ from our Eq.\ (\ref{eq:evo_final}), we
can readily show that they each satisfy Eq. (\ref{eq:ad_evo})
with respect to the cusp anomalous dimension.
We emphasize that the parameter $\eta$ in our approach plays a role
akin to the rapidity parameter $\zeta$ in the additional evolution
equation of Collins and Soper, with Eq.\ (\ref{eq:evo_final}) being
the analogue of the Collins-Soper equation.

Therefore, the dependence on the dimensional regularization scale
$\mu$ of the re-defined TMD PDF (\ref{eq:tmd_re-definition}) is given
by the following renormalization-group equation
\begin{equation}
  \halb \m \frac{d}{d\m}
  \ln f_{q/q}^{\rm mod}(x, \mbox{\boldmath$k_\perp$}; \mu, \eta)
=
  \frac{3}{4} \frac{\a_s}{\pi}\, C_{\rm F} + O(\a_s^2)\ .
\end{equation}
It is important to appreciate that only the modified TMD PDF, given by
Eq.\ (\ref{eq:tmd_re-definition}), obeys such a simple UV-evolution;
without the soft counter term, non-trivial extra contributions would
arise in the corresponding anomalous dimension on the right-hand side
of Eq.\ (\ref{eq:anom-dim}).
Taking logarithmic derivatives of
$f_{q/q}^{\rm mod}(x, \mbox{\boldmath$k_\perp$}; \mu, \eta)$
with respect to both scales $\mu$ and $\eta$, we get
\begin{equation}
  \mu \frac{d}{d\mu}
  \[ \eta \frac{d}{d\eta}
    f_{q/q}^{\rm mod}(x, \mbox{\boldmath$k_\perp$}; \mu, \eta)
  \]
=
  0 \ ,
\label{eq:cons}
\end{equation}
which establishes the formal analogy between our approach and the
Collins-Soper one.
This equation ensures the absence of extra UV-singularities related
to artefacts owing to the light-cone gauge and is equivalent to our
initial requirement of the cancellation of undesirable divergences.

(ii) \emph{Integrated modified PDF.}
Consider now the integration of Eq.\ (\ref{eq:tmd_re-definition})
over $\vecc k_\perp$.
We collect all $\eta$-dependent terms from the virtual and the
real-gluon contributions---contributing UV divergences---and perform
the $\vecc k_\perp$ integration.
We find that the result is $\eta$-independent, so that the DGLAP
evolution of this quantity is guaranteed.
Below, we demonstrate this cancellation explicitly.
The integration of the UV-divergent term (\ref{eq:gamma_1}) trivially
gives
\begin{equation}
  2\frac{\alpha_s}{\pi}\,
  C_{\rm F}\, \frac{1}{\epsilon} \ln \bar \eta \delta(1-x)
  \int\! d^{2} \vecc k_\perp \delta^{(2)} (\vecc k_\perp)
=
  2\frac{\alpha_s}{\pi}\, C_{\rm F}\, \frac{1}{\epsilon}
  \ln \frac{\eta}{p^+} \delta(1-x) \ .
\label{eq:int_uv1}
\end{equation}
On the other hand, the integration (in the dimensional regularization)
of the $\eta$-dependent real-gluon contribution (\ref{eq:real_a})
yields
\begin{eqnarray}
  - \frac{\alpha_s}{\pi^2}\,
    C_{\rm F}\, \ln \frac{\eta}{p^+}
    \delta(1-x) \mu^{2\epsilon}
    \int\! d^{2- 2\epsilon} \vecc k_\perp
    \frac{1}{\vecc k_\perp^2 + \Lambda^2 }
\!\!\! && =
    - 2\frac{\alpha_s}{\pi}\, C_{\rm F}\,
    \Gamma (\epsilon) \ \ln \frac{\eta}{p^+}
\nonumber \\
&& ~~\times
    \delta(1-x)\(\frac{4\pi \mu^2}{\Lambda^2} \)^\epsilon \ ,
\label{eq:int_uv2}
\end{eqnarray}
where $\Lambda^2 = x \lambda^2 - x (1-x) p^2$.
After extracting the UV-divergent term
\begin{equation}
    - 2\frac{\alpha_s}{\pi}
    C_{\rm F}\ \frac{1}{\epsilon} \
    \ln \frac{\eta}{p^+} \delta(1-x)  \ ,
\label{eq:int_uv3}
\end{equation}
we observe that it exactly cancels the right-hand-side of
Eq.\ (\ref{eq:int_uv1}).
The same cancellation occurs between the $\eta$-dependent terms in the
virtual and the real-gluon contributions of the soft factor.
Integration over the transverse momentum (using dimensional
regularization) in the modified TMD PDF (\ref{eq:tmd_re-definition})
yields---at least formally---the integrated PDF
\begin{equation}
   \int\! d^{\omega-2} \vecc k_\perp
   f_{i/a}^{\rm mod} (x, \vecc k_\perp ; \mu , \eta)
   =
   f_{i/a} (x, \mu)
\label{eq:int_form}
\end{equation}
which bears no $\eta$-dependence as well.

From the above considerations it becomes apparent that the
renormalization-group properties of this distribution are
described by the DGLAP equation
\begin{equation}
  \mu \frac{d}{d\mu} f_{i/a} (x, \mu)
=
  \sum_j \int_x^1\! \frac{dz}{z} \ P_{ij}
  \(\frac{x}{z}\) f_{j/a} (z, \mu) \ ,
\label{eq:dglap}
\end{equation}
where the integral kernel reads (in leading order)
\begin{equation}
   P_{ij} (x )
=
   \frac{\alpha_s}{\pi} C_{\rm F} \
   \[ \frac{3}{2} \delta(1-x) + \frac{1+x^2}{(1-x)_+} \]
   + O(\alpha_s^2)
\label{eq:kernel}
\end{equation}
and the $()_+$-regularization is defined in the standard manner by
\begin{equation}
   \int_0^1\! dz \ \frac{f(z)}{(1-z)_+}
   \equiv
   \int_0^1\! dz \ \frac{f(z) - f(0)}{1-z} \ .
\label{eq:auxiliary}
\end{equation}
Thus, one may conclude that the extra cusp-dependent terms are not
present in the integrated case and, consequently, the UV properties of
the standard PDFs (governed by the DGLAP equation) are not affected
by the additional parameter $\eta$, as expected.

\section{Drell-Yan and universality}
\label{sec:DY-univ}
The study presented in the preceding sections was performed for the
semi-inclusive DIS (SIDIS).
In this section we discuss the DY case within our approach and
comment on universality.

We now ask ourselves to what extent our analysis can be applied
to other reactions, like the DY lepton-pair production.
It was shown by Collins \cite{Col02} that the direction of the
integration contours in the lightlike gauge links entering the
definition of the TMD PDF should be reversed relative to the SIDIS,
see Fig.\ \ref{fig:st_dy}:
\begin{eqnarray}
&& [\xi^-, \mbox{\boldmath$\xi_\perp$};
   \infty^-, \mbox{\boldmath$\xi_\perp$}]^\dagger
   [\infty^-, \mbox{\boldmath$0_\perp$};
   \xi^-, \mbox{\boldmath$0_\perp$}]_{_{\rm SIDIS}}
\longrightarrow \\
   && [\xi^-, \mbox{\boldmath$\xi_\perp$};
   - \infty^-, \mbox{\boldmath$\xi_\perp$}]
   [- \infty^-, \mbox{\boldmath$0_\perp$};
   0^-, \mbox{\boldmath$0_\perp$}]_{_{\rm DY}}^\dagger \ .
\nonumber
\label{eq:sidis_dy_col}
\end{eqnarray}
Later, Belitsky-Ji-Yuan \cite{BJY02} argued that the transverse
gauge links, they had introduced to exhaust the gauge invariance
in the light-cone gauge, should be used in the DY case with the
reverse sign, keeping in mind that the lightlike gauge links do not
contribute.
Hence, one has for DY the following combination of gauge links
\begin{eqnarray}
   && [\infty^-, \mbox{\boldmath$\xi_\perp$};
   \infty^-, \mbox{\boldmath$\infty_\perp$}]^\dagger
   [\infty^-, \mbox{\boldmath$\infty_\perp$};
   \infty^-, \mbox{\boldmath$0_\perp$}]_{_{\rm SIDIS}}
\longrightarrow \\
   && [- \infty^-, \mbox{\boldmath$\xi_\perp$};
   - \infty^-, \mbox{\boldmath$\infty_\perp$}]
   [- \infty^-, \mbox{\boldmath$\infty_\perp$};
   - \infty^-, \mbox{\boldmath$0_\perp$}]_{_{\rm DY}}^\dagger \ .
\nonumber
\label{eq:sidis_dy_bjy}
\end{eqnarray}

\begin{figure}[t]
 $$\includegraphics[angle=90,width=0.9\textwidth]{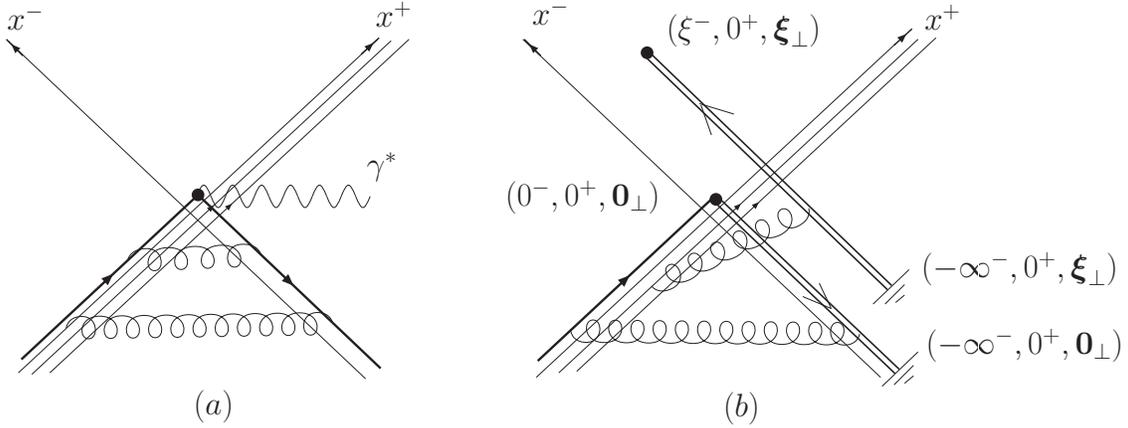}$$
   \vspace{0.0cm} \caption{(a) Spacetime picture of the Drell-Yan
                           process at the amplitude level.
                           The thick line denotes the struck quark,
                           whereas the other solid lines along the
                           $x^+$ direction represent spectators, and
                           curly lines mark exchanged gluons attaching
                           to the gauge links (double lines).
                           (b) Splitting the gauge link in the DY
                           process in analogy to SIDIS in Fig.\
                           \ref{fig:DIS-split} and using the same
                           designations as there.
                           Notice that the two gauge links are
                           separated by a transverse distance
                           $\vecc \xi_\perp$ off the light
                           cone.
\label{fig:st_dy}}
\end{figure}

In our approach, the latter replacement should be supplied with a
change of sign of the additional regulator $\eta$ (cf.\ the pole
prescription in Eq.\ (\ref{eq:pole-prescription})), i.e.,
\begin{eqnarray}
&&   \eta_{_{\rm SIDIS}}
\longrightarrow
   - \eta_{_{\rm DY}}  \ .
\label{eq:sidis_dy_our}
\end{eqnarray}
This change reflects, in fact, the different behavior of the gauge
fields at the plus and minus light-cone infinity subject to the proviso
of different pole-prescriptions.
In the non-polarized case (we exclusively discuss in the present
investigation), this affects only the purely imaginary terms of the
UV-divergent parts---consult Eq.\ (\ref{eq:s_div1}).
Thus, in the DY case, one has
\begin{equation}
  {\S}_{\rm UV}^{(a)}\Big|_{\rm DY}
=
 -\frac{\a_s}{4\pi} C_{\rm F}\frac{1}{\e}
  \[1  - \ln 4 \pi + \gamma_E  - \frac{2\gamma^+ \hat p}{p^+}
  \( 1 + \ln \frac{\eta}{p^+} + \frac{i\pi}{2}
       + i \pi \ C_\infty^{\rm DY} \)
  \] \ ,
\label{eq:s_div_dy}
\end{equation}
where the numerical factor $C_\infty^{\rm DY}$ differs from the SIDIS
case and is defined as
\begin{equation}
  C_\infty^{\rm DY}
=
  \left\{
  \begin{array}{ll}
  &  -1  \ , \ {\rm Advanced}   \\
  & \ \ 0 \ , \ {\rm Retarded}  \\
  & \ \ \frac{1}{2} \ , \ {\rm Principal~~Value~.}
  \
  \end{array} \right.
\label{eq:c_inf_dy}
\end{equation}
These imaginary terms occur at the intermediate steps of the
calculations, but do not contribute to the final expressions for
the unpolarized TMD PDFs.
Exactly the same arguments hold for the additional soft factor
(\ref{eq:soft-factor1}).
This means that the (real-valued) UV anomalous dimension of the
unpolarized TMD PDFs is universal as regards the SIDIS and the DY
processes:
\begin{equation}
   \gamma_{f_{q/q}}^{\rm SIDIS}
   =
   \gamma_{f_{q/q}}^{\rm DY} \ .
\end{equation}
This, however, may not be true for the spin-dependent TMD PDFs,
since in that case the imaginary parts play a crucial role and, thus,
a sign change (expressed in (\ref{eq:s_div_dy})) might indeed affect
the renormalization-group properties and the corresponding evolution
equations.
This is an interesting task which will be pursued separately
elsewhere.

\section{Summary and Conclusions}
\label{sec:concl}

In this paper we have applied renormalization-group techniques to
TMD PDFs, defined in a gauge-invariant way.
We have shown by explicit calculation in the light-cone gauge of
the one-loop gluon radiative corrections to the quantity
$f_{q/q}\left(x,\vecc k_\perp\right)$
that a contribution appears, which is proportional to $\ln p^+$.
This contribution gives rise to an anomalous dimension that is
formally equal to the universal cusp anomalous dimension and helps
unravel a cusp obstruction in the composed gauge contour at
light-cone infinity.
The origin of this anomalous dimension can be traced to the
renormalization effect on the junction point of the split contours,
each associated with a gauge link and attached individually to a
quark field---transforming it into a Mandelstam path-dependent field
\cite{Man62,Man68YM}.
Guided by this finding, we derived a generalized factorization rule
for cusp-connected gauge links which contains an additional phase
factor and worked it out.
For gauge links joined along lightlike contours, this factor reduces
to unity, while for more convoluted contours which run off to
light-cone infinity in the transverse configuration space, this
eikonal factor contributes an anomalous dimension due to the cusp.
In this context, we emphasize that we verified that integrating over
the transverse momenta in
$f_{q/q}\left(x,\vecc k_\perp\right)$
no artefact owing to the contour cusp remains, thus invigorating
the validity of the standard integrated PDF.

In order to eliminate the cusp anomalous dimension and recover the
well-known results in a covariant gauge, where the gauge field
vanishes at infinity, we proposed a new definition for the TMD PDF,
which includes a soft counter term in the sense of Collins and
Hautmann \cite{CH99,CH00}, in order to eliminate the contribution
from the cusp anomalous dimension.
This counter term enters in addition to the transverse gauge inks,
previously introduced by Belitsky, Ji, and Yuan \cite{BJY02}, and
comprises two eikonal factors caused by a particle-like current
flowing along a cusped contour meandering from
$\left(0^-,-\infty^+,\vecc 0_{\perp}\right)$ to
$\left(\xi^-,\infty^+,\vecc \xi_{\perp}\right)$
with a sharp bend in the transverse direction.
We have argued that each of these eikonal factors resembles in
crucial aspects the ``intrinsic Coulomb phase'' found before by
Jakob and Stefanis \cite{JS91} in a formulation of QED in terms
of Mandelstam fields.
In the present case, the cusp-like junction point of the two
individual gauge contours plays a similar role as the so-called
``particle behind the moon'', postulated in QED in connection with the
Lorentz-covariance restoration of its charged sector.
Both quantities share the feature of being ``hidden'' at infinity
and reveal themselves only in terms of (path-dependent) phases,
being independent of external charge distributions (QED case)
and unrelated to boundary conditions to avert light-cone
singularities (TMD PDF case in QCD).
The origin of the phenomenon is in both cases the same and peculiar
to the inclusion of the individual path-dependent exponential into
the field operator supposed to describe the quark as a Mandelstam
field.
No such effect appears in cases where the dynamics of the process
allow one to use a direct contour between the two field points.
In that case, one has to deal only with the connector which has
well-known renormalization properties \cite{CS81,Ste83}.

The ``intrinsic Coulomb phase'' in QED tells us that each charged
particle, though primordially separated from its balancing
counterpart, is still in harness with it.
In the TMD PDF case, this phase accumulates effects due to the
interaction of the struck quark with its target spectators, as
pointed out by Ji and Yuan in \cite{JY02} and reinforced by
Belitsky, Ji, and Yuan in \cite{BJY02}.
However, the existence of a cusp at light-cone infinity went
unnoticed, because in previous works the UV divergences of the TMD
PDF were not considered.
In \cite{JMY04} UV divergences were addressed, but only within the
Collins-Soper approach, which is formulated off-the-light-cone and,
hence, the $\ln p^+$ term does not appear there at all.

The appearance of the cusp anomalous dimension in the present context
is, in actual fact, not really surprising.
We know from the so-called modified factorization of exclusive
reactions that retaining transverse degrees of freedom amounts to the
inclusion of Sudakov factors for each quark in the hard-scattering
subprocess \cite{LS92}---see for a review \cite{Ste99}.
The connection of the Sudakov factors to the cusp anomalous dimension
within the modified factorization scheme was worked out in detail in
\cite{SSK00} up to the level of the next-to-leading-order logarithmic
accuracy.

Our results may have a wide range of applications.
Chief among them:
\begin{itemize}
\item More precise data analyses of various experimental data on
      hard-scattering cross sections.
\item Development of more accurate Monte-Carlo event generators (to
      estimate exclusive components of inclusive cross sections)
      \cite{Col03,HJ07}.
\item Better description of polarized TMD PDFs and the phenomenology
      related to SSA and spin physics
      \cite{Siv89,Siv90,Col02,AM98,MT95,DM07}.
\end{itemize}

\acknowledgments
We would like to thank A.\ P.\ Bakulev, A.\ V.\ Efremov,
A.\ I.\ Karanikas, S.\ V.\ Mikhailov, P.\ V. \ Pobylitsa, and
O.\ V.\ Teryaev for stimulating discussions and useful remarks.
This investigation was partially supported by the Heisenberg-Landau
Programme (grants 2007 and 2008), the Deutsche
Forschungsgemeinschaft under contract 436RUS113/881/0, and the
Russian Federation President's Grant
1450-2003-2.

\end{document}